\documentclass[11pt]{article}
\usepackage{amssymb}

\makeatletter
\def\numberbysection{\@addtoreset{equation}{section}
 \def\theequation{\thesection.\arabic{equation}}}
\makeatother

\newcommand{\be}{\begin{eqnarray}}
\newcommand{\ee}{\end{eqnarray}}
\newcommand{\non}{\nonumber}
\newcommand{\tr}{\mathop{\rm tr}\nolimits}
\def\q{\vartheta}  

\begin{document}
December 2003  \strut\hfill hep-th/0312012 \\
\strut\hfill LAPTH-1008/03 \\ 
\strut\hfill SFB288-603 \\
\vspace*{5mm}

\begin{center}
{\Large  Thermodynamics and conformal properties \\ [1mm]
 of XXZ chains with alternating spins} \\[8mm]

  {\sc Andrei G. Bytsko} $^{a,b}$ and 
  {\sc Anastasia Doikou} $^{c}$

\normalsize{$^{a}$ Steklov Mathematics Institute, Fontanka 27, \\ 
191023 St. Petersburg, Russia \\ [0.5mm]
$^{b}$ Institut f\"ur Theoretische Physik,
 Freie Universit\"at Berlin\\
 Arnimallee 14, 14195 Berlin, Germany \\ [0.5mm]
$^{c}$ Theoretical Physics Laboratory 
  of Annecy--Le--Vieux,\\
LAPTH, B.P. 110, Annecy--Le--Vieux, F-74941, France  }
\footnotetext{e-mail: bytsko@pdmi.ras.ru, \ 
              doikou@lapp.in2p3.fr} 
\end{center}
\vspace{1mm}

\begin{abstract}

The quantum periodic XXZ chain with alternating spins is studied.
The properties of the related R-matrix and Hamiltonians
are discussed. A compact expression for the ground state 
energy is obtained. 
The corresponding conformal anomaly is found via the 
finite-size computations and also by means of the Bethe 
ansatz method. In the presence of an external magnetic 
field, the magnetic susceptibility is derived. 
The results are also generalized to the case
of a chain containing $l$ different spins.

\end{abstract}

\section{Introduction}

($1{+}1$)--dimensional integrable -- exactly solvable -- 
field theories and one dimensional integrable lattice models 
have been extensively studied during the last two decades. 
They found numerous applications in a variety of areas in 
statistical physics \cite{bax,kib}, condensed matter physics 
\cite{ftw}, and high energy physics~\cite{fakor}.  
Apart from that, integrable models have their own mathematical 
interest in the context of quantum groups \cite{Dri,Jim}.

The most known examples of quantum integrable models are 
the sine--Gordon model for which physical quantities such as 
spectrum and exact scattering matrices have been determined 
\cite{zamo}, and the Heisenberg model solved originally by Bethe 
\cite{bethe}. For the Heisenberg model as well, quantities 
of physical interest, such as spectrum, scattering matrices, 
and correlation functions have been explicitly derived
\cite{kib,FT2,korepin}. 

An integrable generalization of the Heisenberg model to 
higher spins and its $q$-deformation (the XXZ model) have 
been extensively studied in the literature 
(see e.g., 
\cite{KuR}--\cite{FY}).
In this paper we will investigate the integrable alternating 
XXZ spin chain, i.e., an inhomogeneous chain with spin $S_1$ at 
odd sites and spin $S_2$ at even sites. The first study of 
such a model, with $(S_1,S_2)=(\frac{1}{2},1)$ was done
in \cite{VEWO}. Later, various aspects of alternating
chains were considered also in 
\cite{AM}--\cite{bado} 
but mostly for specific cases like $(\frac{1}{2},1)$,
$(\frac{1}{2},S)$, and $(S,S{+}\frac{1}{2})$. Our aim
is to elaborate the generic $(S_1,S_2)$ case in detail
in the framework of the algebraic Bethe ansatz with a 
particular emphasis on the thermodynamic limit and the 
finite size correction analysis and then to generalize
the results to the case of $l$ spins. 
In the present paper we will 
consider only the case for periodic boundary conditions.
Some related results on open chains were presented 
in~\cite{doikou2}.

The paper is organized as follows. In Section 2 we outline
the basic ingredients of the algebraic Bethe ansatz approach
for the model in question, in particular, we discuss the
structure and properties of the related R-matrix
and Hamiltonians (technical details are given in the
Appendix~A). In Section 3 we consider the model at
zero temperature. We show that its ground state energy
can be expressed in terms of a specific $q$-deformation
of the $\Psi$-function (related to the Barnes double
gamma function). Here we also present some finite-size
computations of the corresponding central charge.
In Section 4 we first derive the Bethe equations
in the thermodynamic limit at finite temperature $T$
and in the presence of a magnetic field~$H$. 
Then we compute the resulting magnetic 
susceptibility in the $T\to 0$ limit. We also derive
the effective conformal anomaly of the system by means of 
the thermodynamic limit of the algebraic Bethe ansatz
and remark that the underlying conformal theory should
have a spinon basis related to the $D_n$ algebra.
In Section 5 we generalize our results to the case
of the XXZ chain containing $l$ different spins.

For compactness of notations, we will use in the text
both the deformation parameter~$\gamma$ 
and~$q \equiv e^{{\rm i} \gamma}$. In Section 2
we assume that $q$ is either real or takes values on the
unit circle, whereas the results of Sections 3, 4, and 5 
are valid only for the latter case. 

\section{Algebraic Bethe ansatz}

\subsection{R-matrix}

The underlying algebra of the XXZ model is the quantum
Lie algebra $U_q(sl_2)$ with the following defining
relations \cite{KuR}
\begin{equation}\label{cS}
  [S^+ , S^- ] = [ 2 S^3 ]_q \,,\qquad
  [S^3 , S^\pm  ] = \pm S^\pm \,,
\end{equation}
where
\be\label{qnum}
 [t]_q = \frac {q^t - q^{-t} }{q- q^{-1} } 
\ee
is the standard definition of q-number. For generic $q$
algebra (\ref{cS}) has the same structure of representations 
as the undeformed algebra $sl_2$. In particular, irreducible 
highest weight representations $V_S$ are parameterized by 
non-negative integer of half-integer spin $S$ and are 
$(2S+1)$-dimensional. In this case,  the generators 
(\ref{cS}) regarded as elements of End(${\mathbb C}^{2S+1}$)
admit the following matrix realization
\be\label{sss}
 (S^+)_{m,n} = (S^-)_{n,m} = \sqrt{[m]_q [2S +1-m]_q} \,
 \delta_{m+1,n} \,, \qquad
 (S^3)_{m,n} = (S+1-m) \, \delta_{m,n} \,,
\ee
where $m,n = 1,...,2S{+}1$.

An L-operator 
associated with the XXZ model can be chosen as follows
\be\label{Lop}
 L^{S}(\lambda) = \left( \begin{array}{cc}  
  [S^3 - {\mathrm i}\lambda ]_q  &  
    q^{-(\frac 12 + {\mathrm i}\lambda)} \, S^- \\
  q^{\frac 12 + {\mathrm i}\lambda} \, S^+  & 
  - [S^3 + {\mathrm i}\lambda ]_q  \end{array} \right) \,.
\ee
The defining relations (\ref{cS}) are then equivalent to
the following matrix exchange relation
\be\label{RLL}
 R_{12}^{\frac 12 \frac 12} (\lambda ) \, 
    L^S_{13}(\lambda + \mu) \, L^S_{23}(\mu) =
    L^S_{23}(\mu) \, L^S_{13}(\lambda+\mu) \, 
 R_{12}^{\frac 12 \frac 12} (\lambda ) \,,
\ee
where 
$ R^{\frac 12 \frac 12} (\lambda ) = 
 L^{\frac 12}(\lambda + \frac{{\mathrm i}}{2} ) $
(up to a normalization, see below).
The lower indices in (\ref{RLL}) specify components of
the tensor product 
$V_{\frac 12} \otimes V_{\frac 12} \otimes V_{S}$
where the \hbox{R-matrix} and the L-operators act.
Actually, the R-matrix in (\ref{RLL}) is just a particular
representation of a more general object, 
\hbox{$R(\lambda) \in [U_q(sl_2)]^{\otimes 2}$},
that satisfies the Yang-Baxter equation
\be\label{YBE}
 R_{12}(\lambda)\, R_{13}(\lambda+\mu) \, R_{23}(\mu) = 
 R_{23}(\mu)\, R_{13}(\lambda+\mu)\, R_{12}(\lambda)\,. 
\ee

For a homogeneous spin chain or a related lattice model,
it suffices to consider the ``diagonal'' solution 
$R^{SS}(\lambda)$ of (\ref{YBE}). However, to deal with
an inhomogeneous chain, it is unavoidable to use the
general solution $R^{S_1 S_2}(\lambda)$. We provide 
the necessary details below.  

Define the ``total spin" operator $\mathbb J$ 
on $V_{S_1} \otimes V_{S_2}$~as 
\be\label{J}
 \mathbb J = \sum_{j= \delta S}^{S_1 + S_2} j {\cal P}_j \,,
\ee
where $\delta S = |S_1-S_2|$, and ${\cal P}_j$ is the projector 
onto $V_j$ in the Clebsch-Gordan decomposition 
of~\hbox{$V_{S_1} \otimes V_{S_2}$}. If $f(x)$ is a function
non-singular on ${\rm spec}\, { \mathbb J }$, then 
eq.~(\ref{J}) allows us to define 
 $f({\mathbb J}) = 
   \sum_{j= \delta S}^{S_1 + S_2} f(j) {\cal P}_j$.
With these notations, a solution 
$R^{S_1 S_2}(\lambda)$ to the $U_q(sl_2)$-Yang-Baxter 
equation (\ref{YBE}) consistent with the L-operator (\ref{Lop}) 
can be written as follows (see Appendix~A)
\begin{eqnarray}
\nonumber 
 R^{S_1 S_2}(\lambda) &=&  
 (-1)^{{\mathbb J} -S_1 -S_2} \,
 q^{\epsilon (S_1(S_1+1) + S_2(S_2+1)
  - {\mathbb J} ({\mathbb J} +1)) } \times \\
 &&  \times \,
 \label{R1}
  \frac{\Gamma_q( \delta S + 1 + {\mathrm i}\lambda)}%
  {\Gamma_q( \delta S + 1 - {\mathrm i}\lambda)} \, 
    \frac{\Gamma_q({\mathbb J} + 1 - {\mathrm i}\lambda)}%
  {\Gamma_q({\mathbb J} + 1 + {\mathrm i}\lambda)} \, 
 \bigl( R^{S_1 S_2}_\epsilon \bigr)^{-1}  \,, 
 \qquad \epsilon =\pm  \,.
\end{eqnarray}
Here $\Gamma_q(x)$ is a q-deformation of the gamma function 
such that \footnote{It should be stressed here that 
eq.~(\ref{Ga0}) alone does not define the q-gamma function 
uniquely. However, we need to deal only with expressions
of the form $\Gamma_q(x+n)/\Gamma_q(x)$, $n \in {\mathbb N}$
that are well defined.}
\be\label{Ga0}
 \Gamma_q(x+1)=[x]_q \, \Gamma_q(x) \,, 
\ee
and $R_\pm$ are the constant universal R-matrices \cite{Dri,KR2},
\be\label{Rc}
 R_\epsilon = q^{-\epsilon S^3 \otimes S^3} \,
 \sum_{n=0}^{\infty} \frac{q^{ \epsilon \frac{1}{2}(n-n^2)}}%
 {\prod_{k=1}^n [k]_q} \,  \Bigl( \epsilon (q^{-1}-q) \, 
 S^\epsilon \otimes S^{-\epsilon} \Bigr)^n \, 
 q^{-\epsilon S^3 \otimes S^3}  \,.
\ee
The $q=1$ limit of (\ref{R1}) gives the 
well--known $sl_2$--solution constructed in~\cite{KRS}. 

As we discuss in Appendix~A, the following relation holds 
\be\label{JP}
 (-1)^{\mathbb J - 2S} \, 
 q^{\epsilon (2S(S+1) - {\mathbb J}({\mathbb J}+1) )} \,
 (R_\epsilon^{SS})^{-1} = {\mathbb P} \,, 
\ee
where $\mathbb P$ is the permutation in $V_{S} \otimes V_{S}$. 
Identity (\ref{JP}) ensures an important property 
of the solution~(\ref{R1}):
\be\label{RP}
 R^{S S}(0) = {\mathbb P} \,.
\ee
Also, one can show that (\ref{R1}) enjoys the following 
symmetries
\begin{eqnarray}
\label{Rsym}
 & R_{1/q}(\lambda) = 
   \bigl( R_q(-\lambda) \bigr)^{-1} =
  \bigl( R_{q}(\lambda) \bigr)^t \,, & \\
\label{R0sym}
 & R_{1/q}(0) = \bigl( R_q(0) \bigr)^{-1} =
 \bigl( R_{q}(0) \bigr)^t = R_{q}(0) \,, &
\end{eqnarray}
where $t$ stands for the matrix transposition consistent 
with~(\ref{sss}).

In order to obtain $R^{S_1 S_2}(\lambda)$ in terms of the 
algebra generators, which is needed, e.g., for constructing 
related Hamiltonians, one can substitute into (\ref{R1}) 
the following explicit form of the projectors 
(see, e.g., \cite{Byt}) 
\begin{equation}\label{Pj}
 {\cal P}_j  = 
 \prod_{ {l=\delta S }\atop{l\neq j} }^{S_1 + S_2} \, 
 \frac{ {\mathbb X}^{S_1 S_2} 
    -[l]_q [l+1]_q}{ [j-l]_q [j+l+1]_q} \,,
\end{equation}
where
\begin{eqnarray}\label{dC2}
\nonumber  &   {\mathbb X}^{S_1 S_2}  =  
  (q^{S^3} \otimes 1) \, \bigl( S^+ \otimes S^-   
 +  S^- \otimes S^+ \bigr) \, (1 \otimes q^{-S^3})  +  
 \frac{\cos\gamma}{2\sin^2\gamma} \, 
 \bigl( 1\otimes 1 + q^{2 S^3} \otimes q^{-2 S^3} \bigr)  & \\
 &  -  \, \frac{1}{2\sin^2\gamma} \,
 \Bigl( (1 \otimes q^{-2 S^3})\, \cos\gamma(2S_1+1) +
 (q^{2 S^3} \otimes 1)\, \cos\gamma(2S_2+1) \Bigr) . &
\end{eqnarray}

Formulae (\ref{R1}), (\ref{Pj}), and (\ref{dC2}) allow us,
in particular, to verify that $R^{\frac{1}{2} S}(\lambda)$
does reproduce the L-operator~(\ref{Lop}). Let us however 
describe another way of doing that. To this end 
we will construct the ``Baxterized'' form of $R(\lambda)$ 
for higher spins. As we show in Appendix~A, 
the projectors admit the following form
\be\label{PRR}
 {\cal P}_j = 
 \prod_{ {l=\delta S} \atop {l\neq j} }^{S_1 + S_2} \, 
  \frac{ q^{ 2 l(l+1) } - 
 q^{ 2 S_1(S_1+1) + 2 S_2(S_2+1) }  R_{+}^{-1} R_- }%
 {  q^{ 2 l(l+1) } - q^{ 2 j(j+1) } } \,,
\ee
This formula along with (\ref{R1}) allows us to find 
an expression for $R(\lambda)$ in terms 
of $R_+$ and~$R_-$ :
\begin{eqnarray}
\label{R12S}
  R^{\frac{1}{2} S} (\lambda) &=&
 \frac{(q-q^{-1})^{-1}}{ [S+\frac{1}{2}+{\mathrm i}\lambda]_q } \,
 \Bigl[ q^{\frac{1}{2} -{\mathrm i}\lambda} 
 \bigl( R_{+}^{\frac{1}{2} S} \bigr)^{-1} -
 q^{ {\mathrm i}\lambda -\frac{1}{2} } 
 \bigl( R_{-}^{\frac{1}{2} S} \bigr)^{-1} \Bigr] ,  \\
\nonumber
  R^{1 S} (\lambda)  &=&  \frac{(q-q^{-1})^{-2}}%
 { [S+ {\mathrm i}\lambda]_q [S+1+ {\mathrm i}\lambda]_q } \,
 \Bigl[ \Bigr. \Bigl( q^{2{\mathrm i}\lambda -1} - q^{-2} \, \varpi \Bigr) 
 \bigl( R_{-}^{1 S} \bigr)^{-1} + 
 q^{4} \, \varpi \, \bigl( R_{+}^{1 S} \bigr)^{-1}  R_{-}^{1 S} 
 \bigl( R_{+}^{1 S} \bigr)^{-1}  \\ 
 && + \Bigl( q^{1-2{\mathrm i}\lambda} + (1- q^2) \, \varpi -
 q^2 \, \varpi^{-1}  \Bigr) 
 \bigl( R_{+}^{1 S} \bigr)^{-1} \Bigl. \Bigr] , 
 \qquad \varpi \equiv \frac{[2S+1]_q}{[4S+2]_q} ,
\end{eqnarray}
etc. {}From (\ref{R12S}) one easily verifies that 
$[S+1+ {\mathrm i}\lambda]_q 
 R^{\frac{1}{2} S} (\lambda - \frac{{\mathrm i}}{2})$
coincides with $L^S(\lambda)$ given by~(\ref{Lop}).

Let us remark that, twisting (\ref{Lop}), we can obtain 
other L-operators related to the XXZ chain. In particular, 
the following one is often used in the literature:
\be\label{Lh}
 \hat{L} (\lambda) = 
 e^{ \gamma (\lambda - \frac{{\mathrm i}}{2}) S^3 } \,
 L (\lambda) \, e^{ \gamma (\frac{{\mathrm i}}{2} -\lambda) S^3 }
 = \left( \begin{array}{cc}  
  [S^3 - {\mathrm i}\lambda ]_q  &  S^- \\  S^+  & 
  - [S^3 + {\mathrm i}\lambda ]_q  \end{array} \right)
\ee
 {}From the Yang-Baxter equation (\ref{YBE}) and the $U(1)$ 
symmetry of the solution (\ref{R1}),
\be\label{RR3}
  [ R_{ab} (\lambda) , S_a^3 + S_b^3 ] = 0 \,,
\ee
it follows that the universal R-matrix corresponding
to (\ref{Lh}) is related to (\ref{R1}) as
\be\label{Rh}
  \hat{R} (\lambda) = 
 e^{ \gamma \lambda (1\otimes S^3) } \, R(\lambda) \,
 e^{ - \gamma \lambda (1\otimes S^3) } \,.
\ee
It has the following symmetries (see Appendix~A)
\be\label{Rhsym}
 \hat{R}(-\lambda) = 
   \bigl( \hat{R}(\lambda) \bigr)^{-1} \,, \qquad
 \hat{R}_{1/q}(-\lambda) = 
   \hat{R}_q(\lambda)  \,, \qquad
 \bigl( \hat{R}(\lambda) \bigr)^t = \hat{R}(\lambda) \,. 
\ee

\subsection{Hamiltonian}

In order to construct a Hamiltonian for a closed alternating 
spin chain we apply a generalization of the method developed
in \cite{TTF} and consider two transfer matrices:
 \footnote{If we introduce proper inhomogeneities of the 
spectral parameter, we will obtain massless relativistic 
dispersion relations for the particle--like excitations 
of the model \cite{bado}.} 
\be\label{tau}
 \tau^{(i)} (\lambda) = \tr_a T_a^{(i)} (\lambda) =
 \tr_a \Bigl( 
 R_{a,N}^{S_i S_2}(\lambda) \, 
   R_{a,N-1}^{S_i S_1}(\lambda) \ldots
 R_{a,2}^{S_i S_2}(\lambda) \, 
   R_{a,1}^{S_i S_1}(\lambda)  \Bigr) \,,
 \quad   i=1,2 \,,
\ee 
where the auxiliary space (denoted by the subscript $a$)
is $V_{S_1}$ and $V_{S_2}$, respectively. Since the 
monodromy matrices satisfy the Yang-Baxter equation
\be\label{Tyb}
 R^{S_i S_j}_{ab}(\lambda-\mu) \, 
  T_a^{(i)} (\lambda) \, T_b^{(j)} (\mu) =
  T_b^{(j)} (\mu) \, T_a^{(i)} (\lambda) \, 
 R^{S_i S_j}_{ab}(\lambda-\mu) 
\ee
the corresponding transfer matrices commute
\be\label{tauyb}
  [ \tau^{(i)} (\lambda) , \tau^{(j)} (\mu)] = 0 \,.
\ee
Therefore, the first and higher order logarithmic 
derivatives of $\tau^{(i)}(\lambda)$ at $\lambda=0$ 
are quantum integrals of motion. Moreover, thanks to 
the property (\ref{RP}), these integrals  are local, 
according to Theorem~3 in~\cite{Tar}.

Using the property (\ref{RP}) (along with 
$\tr_a {\mathbb P}_{ab}=1$), we find that
\be\label{tau0}
\tau^{(1)} (0) = 
 \Bigl( \prod_{k - {\rm odd}}^{N-3} \! 
     {\mathbb P}_{k,k+2} \Bigr) 
 \Bigl( \prod_{k - {\rm odd}}^{N-1} \! 
     R_{k,k+1}^{S_1 S_2}(0) \Bigr) ,  \qquad
\tau^{(2)} (0) = 
 \Bigl( \prod_{k - {\rm even}}^{N-2} \! 
    {\mathbb P}_{k,k+2} \Bigr) 
 \Bigl( \prod_{k - {\rm even}}^{N} \! 
    R_{k,k+1}^{S_2 S_1}(0) \Bigr) ,
\ee 
where $N+1\equiv 1$ (the periodic boundary conditions).
Noticing that for both $\tau(0)$'s in (\ref{tau0}) 
all $R(0)$'s commute among each over, it is not difficult 
to obtain the following expressions for Hamiltonians
\be\label{H12}
 {\cal H}^{(1)} = \frac{{\mathrm i}}{2} \,
  \partial_\lambda \ln \tau^{(1)}(\lambda) \Bigm|_{\lambda=0} =
 \sum\limits_{n - {\rm odd}}^{N-1} \! H^{(1)}_{n} , \qquad
 {\cal H}^{(2)} = \frac{{\mathrm i}}{2} \,
  \partial_\lambda \ln \tau^{(2)}(\lambda)  \Bigm|_{\lambda=0} = 
 \sum\limits_{n - {\rm even}}^{N} \! H^{(2)}_{n} 
\ee 
with $H_n^{(i)}$ being local Hamiltonians involving 
three nearest sites:
\begin{equation}\label{H1}
  H^{(i)}_{n} = 
  \Bigl( R^{S_i S_{\tilde \imath}}_{n,n+1}(0) \Bigr)^{-1} \, 
 \Bigl[ \frac{{\mathrm i}}{2} \,
 \partial_\lambda R^{S_i S_{\tilde\imath}}_{n,n+1}(\lambda) 
 \Bigm|_{\lambda=0}  +\, h^{S_i S_i}_{n,n+2} \, 
 R^{S_i S_{\tilde  \imath}}_{n,n+1}(0) \Bigr] \,,
\end{equation}
where ${\tilde \imath}=i{+}1\ (\bmod 2)$, and
\be\label{hs}
 h^{S S}_{nm} = \frac{{\mathrm i}}{2} \, {\mathbb P}_{nm} \, 
 \partial_\lambda R^{S S}_{nm}(\lambda) \Bigm|_{\lambda=0} 
\ee
is the local Hamiltonian of the homogeneous spin S chain.
These formulae provide a particular case of local Hamiltonians 
for inhomogeneous chains derived in \cite{Tar} 
(see also \cite{VEWO}). Obviously, they recover the 
homogeneous case as well since for $S_1=S_2$ we have 
$R^{S_1 S_2}_{n,n+1}(0) = {\mathbb P}_{n,n+1}$, and then
$H_n^{(i)}$ becomes a sum of two terms of the type~(\ref{hs}).

Let us notice that, computing in the same way the
local Hamiltonians $\hat{H}^{(i)}_n$ corresponding
to the R-matrix (\ref{Rh}) and taking the property
(\ref{RR3}) into account, we obtain
\be\label{Hh}
 \hat{H}^{(i)}_n = H^{(i)}_n + 
 \frac{{\mathrm i}\gamma}{2} \bigl( S^3_n - S^3_{n+2} \bigr) \,.
\ee
Hence for the total Hamiltonians we have
$\hat{\cal H}^{(i)} = {\cal H}^{(i)}$.

Let $\Psi_q(x)$ denote logarithmic derivative of the 
$q$-gamma function used in (\ref{R1}), so that we have,
in particular, $\Psi_q(x+1)=\Psi_q(x)+ 
 \gamma \cot \gamma x$. Then (\ref{R1}) allows
us to rewrite the local Hamiltonian in the following,
more explicit, form
\begin{eqnarray}
\nonumber   
 && H^{(i)}_{n} = (R_\epsilon)_{n,n+1} \, 
 q^{\epsilon {\mathbb J}_{n,n+1} ({\mathbb J}_{n,n+1} + 1)} \,
 (-1)^{{\mathbb J}_{n,n+1}} \, 
 \Bigl[ \Psi_q({\mathbb J}_{n,n+1} +1) - \Psi_q(\delta S + 1) \\
 && \phantom{ H^{(1)}_{n} =} + 
 \Psi_q({\mathbb J}_{n,n+2} +1) - \Psi_q(1) \Bigr] \, 
 (-1)^{{\mathbb J}_{n,n+1}} \,
 q^{-\epsilon {\mathbb J}_{n,n+1} ({\mathbb J}_{n,n+1} + 1)} \, 
 (R^{-1}_\epsilon)_{n,n+1} \,,
\label{Hxa}  
\end{eqnarray}
where the quantum space is $V_{S_i}$ at the sites $n$, $n{+}2$,
and $V_{S_{\tilde \imath}}$ at the site $n{+}1$.

In order to exemplify (\ref{H1}) and (\ref{Hxa}), let us 
construct $H^{(1)}_{n}$ explicitly for $(S_1,S_2)=(\frac 12,S)$.
Consider the XXX case first. Here we need only the 
following projectors
\be
 \bigl( {\cal P}^{\frac 12  S}_{S \pm \frac 12} \bigr)_{n,m} = 
 \frac{1}{2S+1} \bigl(S + \frac 12 \pm \frac 12 \pm 
 \vec{\sigma}_n \cdot \vec{S}_m \bigr) \,, 
\ee
where $\vec{\sigma}=(\sigma^1,\sigma^2,\sigma^3)$ consists 
of the Pauli matrices. In this case (\ref{Hxa}) yields
\begin{eqnarray}
 \nonumber 
 H^{(1)}_{n} &=&  \frac{1}{(2S+1)^2} \Bigl( 
 2 (\vec{\sigma}_n \cdot \vec{S}_{n+1}) +
 2 (\vec{S}_{n+1} \cdot \vec{\sigma}_{n+2}) \\
 \label{Hx1}
 && + \,  \bigl\{ (\vec{\sigma}_n \cdot \vec{S}_{n+1}) ,
     (\vec{S}_{n+1} \cdot \vec{\sigma}_{n+2}) \bigr\} 
 + \bigl( \frac{1}{4} - S(S+1) \bigr) \,
 (\vec{\sigma}_{n} \cdot \vec{\sigma}_{n+2}) + \frac{3}{4} \Bigr) \,.
\end{eqnarray}
Here and below $\{\,,\,\}$ stands for the anticommutator.
Since 
$\{ (\vec{\sigma} \cdot \vec{a}),(\vec{\sigma} \cdot \vec{b}) \}=
 2 (\vec{a} \cdot \vec{b})$, it is easy to see that, for $S=\frac 12$,
eq.~(\ref{Hx1}) reduces to the usual Heisenberg Hamiltonian.
In a slightly different form (\ref{Hx1}) was given 
also in~\cite{AM}.

In the XXZ case, it is easier to compute $\hat{H}_n^{(1)}$, 
taking into account the symmetry (\ref{Rhsym}). To write down 
the answer in a compact form, we introduce the Sklyanin 
generators \cite{Skl} of~$U_q(sl_2)$,
\be\label{Sg}
 {\mathfrak S}^0 = [\frac 12]_q \, \cos (\gamma S_3) \,, \quad
 {\mathfrak S}^1 = \frac{1}{2} (S^+ + S^-) \,, \quad
 {\mathfrak S}^2 = \frac{1}{2{\mathrm i}} (S^+ - S^-) \,, \quad
 {\mathfrak S}^3 = \cos(\frac{\gamma}{2}) \, [S_3]_q \,. 
\ee
They satisfy the following quadratic relations
\be\label{Sa} 
 [ {\mathfrak S}^3 , S^\pm] = \pm \, \{ {\mathfrak S}^0 , S^\pm \} 
 \,,\qquad
 [ {\mathfrak S}^0 , S^\pm] = \mp \, (\tan \frac{\gamma}{2})^2 \,
  \{ {\mathfrak S}^3 , S^\pm \}  \,.
\ee
Using these relations, we obtain from (\ref{H1})-(\ref{Hh}) 
the following $(\frac 12,S)$--Hamiltonian (see also \cite{VEWO} 
for a $(\frac 12,1)$--Hamiltonian)
\begin{eqnarray}
 \nonumber 
 \hat{H}^{(1)}_{n} &=&  \frac{1}{4[S+\frac 12]_q^2} \Bigl( 
 2 \bigl\{ {\mathfrak S}^0_{n+1} ,
  (\vec{\sigma}^\gamma_n \cdot \vec{\mathfrak S}_{n+1}) +
  (\vec{\mathfrak S}_{n+1} \cdot \vec{\sigma}^\gamma_{n+2}) \bigl\} 
 + \cos(2\gamma S^3_n) \\
 \nonumber
  && + \,  \bigl\{ (\vec{\sigma}_n \cdot \vec{\mathfrak S}_{n+1}) ,
   (\vec{\mathfrak S}_{n+1} \cdot \vec{\sigma}^\gamma_{n+2}) \bigr\} 
 + \bigl( ({\mathfrak S}^0_{n+1})^2 -  
      \vec{\mathfrak S}_{n+1} \cdot \vec{\mathfrak S}_{n+1} \bigr) \,
 (\vec{\sigma}_{n} \cdot \vec{\sigma}^\gamma_{n+2}) \\
 \label{Hz}
 && + \, \frac{3}{4}\cos\gamma +
 \bigl[ {\mathfrak S}^0_{n+1} , 
   [ (\vec{\sigma}_n \cdot \vec{S}_{n+1}), 
    (\vec{S}_{n+1} \cdot \vec{\sigma}^\gamma_{n+2}) ]\bigr]
 \Bigr) \,,
\end{eqnarray}
where $\vec{\mathfrak S} = ({\mathfrak S}^1 , {\mathfrak S}^2 , 
 {\mathfrak S}^3)$ and 
$\vec{\sigma}^\gamma = (\sigma^1, \sigma^2, \sigma^3 \cos\gamma)$.
For $S=\frac 12$ we have ${\mathfrak S}^0 = \frac 12$, 
$\vec{\mathfrak S} = \frac{1}{2} \vec{\sigma}$ and then (\ref{Hz})
reduces to the usual XXZ--deformation of the Heisenberg Hamiltonian.

Consider the $*$--structure on $U_q(sl_2)$ corresponding to the 
compact real form $U_q(su(2))$, i.e., an anti-automorphism such that
\begin{equation}\label{star}
  (S^\pm)^* = S^\mp \,, \qquad  (S^3)^* = S^3 
\end{equation}
and that extends onto a tensor product as 
$(\xi\otimes \zeta)^* = \xi^* \otimes \zeta^*$.
If $q$ is real or $|q|=1$ we have (see Appendix~A)
the following ``unitarity'' relations
\be\label{*R}
 \bigl(R_q(\lambda)\bigr)^* =  
   \bigl( R_{\bar{q}}(\bar{\lambda}) \bigr)^{-1} \,, \qquad
 \bigl(\hat{R}_q(\lambda)\bigr)^* = 
   \bigl( \hat{R}_q(\bar{\lambda}) \bigr)^{-1} \,.
\ee
Applying the latter relation along with (\ref{R0sym}) 
to the counterpart of (\ref{H1}) for $\hat{H}^{(i)}_{n}$
and taking into account that $h_{nm}^*=h_{nm}$ 
(see, e.g.~\cite{Byt}), we infer that 
\be\label{x*}
  \bigl( \hat{H}^{(i)}_{n} \bigr)^* = \hat{H}^{(i)}_{n} \,.
\ee
Therefore, the total Hamiltonian of an alternating spin chain
${\cal H}^{(i)}$ is hermitian if $\gamma \in {\mathbb R}$
or~${\rm i} \gamma \in {\mathbb R}$.
  
\subsection{Momentum operator}
Unlike the homogeneous case, the transfer matrix
$\tau^{(i)}(0)$ is not the shift operator and hence its
logarithm cannot be interpreted as the momentum operator. 
Observe however that the products of permutations
entering (\ref{tau0}),
\be
 U^{(1)} =
 \prod_{k - {\rm odd}}^{N-3}  {\mathbb P}_{k,k+2} \,, \qquad
 U^{(2)} = \prod_{k - {\rm even}}^{N-2}  {\mathbb P}_{k,k+2} \,,
\ee
are shift operators of odd/even lattice sites by two steps, i.e.,
\be\label{Ux}
 U^{(i)} \, \xi_n = \xi_{n+2} \, U^{(i)} \qquad
 {\rm if}\ \  n\equiv i \bmod 2 \,.
\ee
This motivates to consider quantum integrals of motion generated
by $\tau^{1,2}(\lambda)=\tau^{(1)}(\lambda) \tau^{(2)}(\lambda)$. 
Due to (\ref{tauyb}), the corresponding Hamiltonian is given by
\be\label{Ht}
 {\cal H} = {\mathrm i}\frac{J}{4} \,
 \partial_\lambda \ln \tau^{(1,2)}(\lambda) \Bigm|_{\lambda=0}
 = \frac J2 \, \bigl( {\cal H}^{(1)} + {\cal H}^{(2)} \bigr) 
\ee
with ${\cal H}^{(i)}$ defined in (\ref{H12}) and $J>0$ being
a coupling constant.

In order to compute $\ln \tau^{1,2} (0)$ we need a property 
of the R-matrix that follows from evaluating the 
Yang-Baxter equation (\ref{YBE}) in 
\hbox{$V_{S_1} \otimes V_{S_2} \otimes V_{S_1}$} 
at $\mu = - \lambda$.  In view of (\ref{RP}) this leads to
the relation
\be\label{YBa}
 R^{S_1 S_2}_{12}(\lambda)\, R^{S_2 S_1}_{21}(-\lambda) = 
 R^{S_2 S_1}_{23}(-\lambda)\, R^{S_1 S_2}_{32}(\lambda)\,, 
\ee
and we infer, in particular, that
\be\label{RR0}
 R^{S_1 S_2}_{12}(0)\, R^{S_2 S_1}_{21}(0) = 1 \otimes 1 
\ee
if the normalization of the R-matrix was chosen appropriately
(as, e.g., in~(\ref{R1})).

Now, using (\ref{tau0}), (\ref{Ux}), and (\ref{RR0}), we obtain
\be\label{taup}
  \tau^{1,2} (0) = U^{(1)} \, U^{(2)} 
\ee
which generates a shift by two lattice sites. Thus, 
${\cal P}=\frac{1}{2{\mathrm i}}\ln \tau^{1,2}(0)$ can be regarded as 
the momentum operator (as was observed also in~\cite{VEWO}). 
Similarly to the homogeneous spin chain, the pair 
$({\cal H},{\cal P})$ defines in the gapless regime 
($|q|=1$) at zero temperature a conformally invariant system.

\subsection{Bethe equations}
Consider the alternating XXZ chain of even length~$N$.
According to the general scheme of the algebraic Bethe ansatz 
\cite{TTF,Tar} (see also \cite{Fad1} for a review), 
eigenvectors and eigenvalues of the auxiliary transfer matrix
\be\label{tauaux}
 \tau^{(0)} (\lambda) = \tr_a T_a^{(0)} (\lambda) =
 \tr_a \Bigl( L_N^{S_2}(\lambda) \, L_{N-1}^{S_1}(\lambda) 
 \ldots L_2^{S_2}(\lambda) \, L_1^{S_1}(\lambda) \Bigr) \,,
\ee 
where the trace is taken over the auxiliary space~$V_{\frac 12}$,
are parameterized by solutions of the system of equations
\begin{equation}\label{BEz}
 \prod_{j=1}^2
 \left( \frac{ \sinh\gamma(\lambda_\alpha + {\mathrm i}S_j)}%
  { \sinh\gamma(\lambda_\alpha - {\mathrm i}S_j)}  \right)^{\frac N2}
 = - \prod_{\beta=1}^M 
 \frac{ \sinh\gamma(\lambda_\alpha - \lambda_\beta + {\mathrm i})}%
  { \sinh\gamma(\lambda_\alpha - \lambda_\beta - {\mathrm i}) } \,, \qquad
 \alpha =1,\ldots,M,
\end{equation}
where $M$ is the level of a Bethe vector 
$\Psi(\lambda_1,\ldots,\lambda_M)$; it is related to the
total $z$--component of the spin as 
\begin{equation}\label{spin}
 \Bigl( \sum_{n=1}^N S^3_n \Bigr) \, 
    \Psi(\lambda_1,\ldots,\lambda_M)
    = S_z \, \Psi(\lambda_1,\ldots,\lambda_M) \,, 
\end{equation}
where $S_z = ( ( S_1 + S_2 ) N/2 - M )$.
The Yang-Baxter equation
\be\label{TTyb}
 R^{\frac 12 S_j}_{ab}(\lambda-\mu) \, 
  T_a^{(0)} (\lambda) \, T_b^{(j)} (\mu) =
  T_b^{(j)} (\mu) \, T_a^{(0)} (\lambda) \, 
 R^{\frac 12 S_j}_{ab}(\lambda-\mu) \,, \qquad j=1,2
\ee
implies commutativity of the corresponding transfer matrices, 
\be\label{ttauyb}
  [ \tau^{(0)} (\lambda) , \tau^{(j)} (\mu)] = 0 \,.
\ee
Therefore the eigenvectors of $\tau^{(0)} (\lambda)$
are simultaneously eigenvectors for~$\tau^{(j)}(\lambda)$.
Moreover, as follows from the results of \cite{Tar}, 
eigenvalues of the momentum operator ${\cal P}$ and 
the Hamiltonian ${\cal H}$  (in the presence of an
external magnetic field $H$) are given by
\be\label{PEz}
 P(\lambda_1,\ldots,\lambda_M) =
  \sum_{\alpha=1}^{M} \sum_{j=1}^2 
  p(\lambda_\alpha | S_j) \,, \qquad
 E(\lambda_1,\ldots,\lambda_M) = 
  - \frac{J}{2} \sum_{\alpha=1}^M \sum_{j=1}^2 
  p^\prime (\lambda_\alpha | S_j) - H \, S_z\,,
\ee
where we introduced the functions
\begin{eqnarray}
 \label{Pa}
 && p(\lambda| S) = \frac{1}{2{\mathrm i}} 
  \ln \frac{ \sinh\gamma(\lambda - {\mathrm i}S)}%
  { \sinh\gamma(\lambda + {\mathrm i}S)} =
   \arctan \bigl( \tanh(\gamma\lambda) \,
   \cot(\gamma S) \bigr) \,, \\
 && 
 \label{Pb}
 p^\prime (\lambda| S) =
 \partial_{\lambda} p(\lambda| S) = 
 \frac{ \gamma \, \sin(2\gamma S)}%
 {\cosh(2\gamma\lambda) - \cos(2\gamma S)} \,.
\end{eqnarray}
The corresponding values for the alternating XXX chain are obtained
by the $\gamma \rightarrow 0$ limit of (\ref{PEz})-(\ref{Pb}).

\section{Ground state energy and finite size effects at $T=0$}
\subsection{Ground state energy }
Let $\pi/\gamma$ be integer.
In the thermodynamic limit, when $M,N \rightarrow \infty$, 
the string hypothesis holds \cite{gt1}, namely solutions of 
the Bethe equations (\ref{BEz}) form string--like clusters, 
i.e., groups of the type
\be
 \label{STR} 
 \lambda_{\alpha}^{(n,k)} &=&
 \lambda_{\alpha}^n + {{\mathrm i}\over 2}(n+1-2k) \,, 
    \qquad k=1,2,\ldots,n, \\
 \lambda_{\alpha}^{(0)}&=&
   \lambda_{\alpha}^{0}+{\mathrm i}\frac{\pi}{2\gamma},
 \non 
\ee 
where the centers of strings $\lambda_{\alpha}^n$ and 
$\lambda^{0}_{\alpha}$ are real, and $\lambda^{(0)}$ is the negative 
parity string. The number $n$ is called the length of a string.

At the zero temperature the system is in its ground state which,
as we will see below, is a Dirac sea filled by strings of lengths 
$2S_1$ and $2S_2$ (negative parity strings are not present).  
Taking logarithm of (\ref{BEz}) and carrying out 
summation along the strings in the standard way, we obtain a set 
of equations on the centers of these strings:
\begin{equation}\label{be3}
 \frac N2 \biggl( \sum_{m=\frac 12}^{2S_i - \frac 12}
 p(\lambda_\alpha^{(i)} | m) +
 \sum_{m=\delta S + \frac 12}^{S_1+S_2 - \frac 12}
 p(\lambda_\alpha^{(i)} | m) \biggr) = \pi Q^{(i)}_\alpha
 + \sum_{j=1}^2 \sum_{\beta=1}^{M_j}  \Phi_{2S_i,2S_j} 
     (\lambda_\alpha^{(i)}-\lambda_\beta^{(j)}) \,,
\end{equation}
where $\delta S = |S_1-S_2|$ and 
$\lambda^{(i)}_\alpha \equiv \lambda^{2S_i}_\alpha$, $i=1,2$.
The numbers $Q^{(i)}_\alpha$ are distinct integer or half-integer 
numbers such that $|Q^{(i)}_\alpha| \leq \frac N8 - \frac 12$, and
\begin{equation}\label{phi1}
 \Phi_{n,m}(\lambda)=
 \sum_{l={\frac 12} |n-m|}^{ {\frac 12}(n+m)-1}
  \Bigl( p( \lambda | l) + p( \lambda | l+1) \Bigr)
\end{equation}
(for $n=m$ the term with $l=0$ is omitted).

When $N,M \rightarrow \infty$, (\ref{be3}) become integral
equations:
\begin{equation}\label{gsii1}
 \sum_{m=\frac 12}^{2S_i - \frac 12} 
 p^\prime(\lambda | m) +
 \sum_{m=\delta S + \frac 12}^{S_1+S_2 - \frac 12}
 p^\prime(\lambda | m)
 = 2 \pi \rho^{(i)} (\lambda) + 2 \sum_{j=1}^2 
 ( \Phi^\prime_{2S_i,2S_j} * \rho^{(j)}) (\lambda) \,,
\end{equation}
where the convolution is defined as 
$(u*v)(\lambda)=\int_{-\infty}^{\infty} d\mu \, u(\lambda-\mu) v (\mu)$, 
and $\Phi^\prime_{n,m}(\lambda)$ is the derivative of (\ref{phi1}).
It is straightforward to check that
\begin{equation}\label{F2}
 \int_{-\infty}^\infty d\lambda \, e^{{\mathrm i}k \lambda} \,
 p^\prime(\lambda | m) = 
 \pi \frac{ \sinh (\frac{k\pi}{2\gamma} - mk) }%
 {\sinh \frac{k\pi}{2\gamma} }
 \qquad {\rm if} \quad  m< \frac \pi\gamma \,.
\end{equation}
Using this and taking Fourier transform of (\ref{gsii1}), it is 
not difficult to see that the only solution of these equations is 
given by
\begin{equation}\label{gsd1}
 \rho^{(1)} = \rho^{(2)} = \frac{1}{4\cosh \pi \lambda} 
 \qquad {\rm if} \quad  2 \gamma \max (S_1,S_2)  < \pi \,.
\end{equation}

As follows from (\ref{PEz}), (\ref{Pb}), and (\ref{gsd1}), 
computation of the ground state energy amounts to 
computing a sum of integrals of the following type
\begin{equation}\label{Igm1}
 I_\gamma(m)= \int_{-\infty}^\infty d\lambda \,
 \frac{p^\prime(\lambda | m)}{\cosh \pi \lambda}  =
 \int_{0}^\infty dk \,
 \frac{1}{\cosh \frac k2 } \,
 \frac{ \sinh (\frac{k\pi}{2\gamma} - mk) }%
 {\sinh \frac{k\pi}{2\gamma} } \,,
\end{equation}
where, in the last equality, we took (\ref{F2}) into account.
Expanding the functions in the denominator in series, we obtain
\begin{eqnarray}
 \nonumber 
 I_\gamma(m) &=&  
 \Psi \Bigl(\frac m2 + \frac 34 \Bigr) -
 \Psi \Bigl(\frac m2 + \frac 14 \Bigr) +
 \sum_{n\geq 1} \Bigl[ \Psi \Bigl(\frac m2 + \frac 34 +
 \frac{\pi n}{2\gamma} \Bigr) -
 \Psi \Bigl(\frac m2 + \frac 14 + \frac{\pi n}{2\gamma} \Bigr)  \\
 &&  \label{Igm3b}
 - \ \Psi \Bigl( -\frac m2 + \frac 34 + \frac{\pi n}{2\gamma} \Bigr) +
 \Psi \Bigl(-\frac m2 + \frac 14 + \frac{\pi n}{2\gamma} \Bigr) \Bigr]\,.
\end{eqnarray}

Let us now show that $I_\gamma(m)$ admits, for real $\gamma$, 
a compact form in terms of a certain deformation of the 
$\Psi$--function. 
To this end we define the following special function
\begin{equation}\label{qgamma1}
 \Gamma_q(x) = (q-q^{-1})^{-x} \,
 \frac{\Gamma_2(bx|b,b^{-1})}{\Gamma_2(b+b^{-1}-bx|b,b^{-1})}  \,,
\end{equation}
where $q=e^{{\mathrm i}\pi b^2}$, $0{<}b{<}1$ and 
$\Gamma_2(x|\omega_1,\omega_2)$ 
is the double gamma function introduced by Barnes \cite{Bar}. 
Properties of $\Gamma_2(x|\omega_1,\omega_2)$
imply the following relations
\begin{equation}\label{qgamma2}
 \Gamma_q(x+1) = [x]_q \, \Gamma_q(x) \,, \qquad
 (q-q^{-1})^{1/b^2} \, \Gamma_q(x+1/b^2) =
 2 \sin(\pi x) \, \Gamma_q(x) \,,
\end{equation}
first of which allows us to regard $\Gamma_q(x)$ as
a $q$--deformation of the gamma function. $\Gamma_q(x)$ is
closely related to the noncompact quantum dilogarithm that
has been actively studied recently~\cite{qd} in the context
of integrable models. In particular, it can be shown that
\begin{equation}\label{qgamma3}
 \ln \Gamma_q(x) =  {\rm const}
  -x \ln (q-q^{-1}) +  \frac{\pi {\mathrm i}}{2} (b+b^{-1}-bx)bx
  - \int\limits_{{\mathbb R}+{\mathrm i}0} \frac{dt}{t} \,
  \frac{e^{btx}}{(1-e^{bt})(1-e^{t/b})}  \,.
\end{equation}
This expression allows us to establish connection between
the integral in (\ref{Igm1}) and the $q$--deformation of the
$\Psi$--function defined via
$\Psi_q(x)\equiv\partial_x \ln \Gamma_q(x)$. Namely, we find that
\begin{equation}\label{qpsi2}
 I_\gamma(m) =
  \Psi_{q^2} \Bigl( \frac m2  + \frac 34 \Bigr) -
  \Psi_{q^2} \Bigl( \frac m2 + \frac 14\Bigr) \,,
  \qquad {\rm where} \ \ q = e^{{\mathrm i}\gamma} 
\end{equation}

Now, from (\ref{PEz}), (\ref{gsd1}), (\ref{Igm1}), and 
(\ref{qpsi2}), we obtain the ground state energy per lattice site
\begin{eqnarray}
 \label{gse1z} 
 & \displaystyle
 e(S_1,S_2) = \frac{E}{N} = - \frac J8  \sum_{i=1}^2
 \Bigl( \sum_{m=\frac 12}^{2S_i - \frac 12} I_\gamma(m) +
  \sum_{m=\delta S + \frac 12}^{S_1+S_2 - \frac 12}
   I_\gamma(m) \Bigr)  & \\
 \nonumber  
 & \displaystyle
 = \frac{J}{4} \left[
 \Psi_{q^2} \Bigl( \frac 12 \delta S + \frac 12 \Bigr) +
 \Psi_{q^2} \Bigl( \frac 12 \Bigr) -
 \Psi_{q^2} \Bigl( \frac 12 (S_1+S_2) + \frac 12 \Bigr) \right]
 - \frac{J}{8} \left[ \Psi_{q^2} \Bigl(S_1 + \frac 12 \Bigr) +
 \Psi_{q^2} \Bigl(S_2 + \frac 12 \Bigr) \right] . &
\end{eqnarray}
For a homogeneous chain, (\ref{gse1z}) reduces to
\begin{equation}\label{Eg3}
 e(S) =  \frac J2 \Bigl[ \Psi_{q^2}\Bigl(\frac 12 \Bigr) -
 \Psi_{q^2}\Bigl(S + \frac 12 \Bigr) \Bigr] \,.
\end{equation}
Replacing $q$-deformed $\Psi$--functions with
ordinary $\Psi$--functions in (\ref{gse1z})-(\ref{Eg3}),
one recovers the energy densities in the XXX case.
In particular, (\ref{Eg3}) turns in this limit into 
the well-known formula~\cite{LT}.

It is worth noticing that the first relation in (\ref{qgamma2})
implies that $\Psi_{q^2}(x+1)-\Psi_{q^2}(x)= 2\gamma \cot 2\gamma x$.
Therefore, if $S_1$ and $S_2$ or, respectively, $S$ are integer, we 
can rewrite (\ref{gse1z}) and (\ref{Eg3}) in terms of 
trigonometric functions:
\begin{eqnarray}
 \label{Eg3'}
 & e(S_1,S_2) = - { \displaystyle \frac{\gamma J}{4}}  \Bigl(
   \sum\limits_{k=0}^{S_1 -1} \cot [\gamma(2k+1)] +
   \sum\limits_{k=0}^{S_2 -1} \cot [\gamma(2k+1)]
    + 2 \sum\limits_{k= \frac 12 \delta S}^{ \frac 12 (S_1+S_2) -1} 
    \cot [\gamma(2k+1)] \Bigr) , &\\
 & e(S) = - \gamma J \sum\limits_{k=0}^{S-1} \cot [\gamma(2k+1)] 
  \,. &
\end{eqnarray}

\subsection{Finite size corrections}
It is well known that statistical systems at criticality  
are expected to exhibit conformal invariance~\cite{bpz}.
Therefore, the critical behavior of such systems should be 
described by a certain conformal field theory. One can identify 
the corresponding effective central charge $c$ by investigating 
the finite size effects of the ground state of a critical 
system~\cite{bc}. For a spin chain of a finite length $N$, 
the ground state energy at zero temperature 
depends on~$N$ as follows:
 \begin{equation}\label{fsc}
 e_N = e_\infty - \frac{\pi c v_s}{6 N^2} + o(N^{-2}) \,,
\end{equation}
where $v_s$ is the speed of sound. With the energy and momentum
normalization as in (\ref{PEz}) we have $v_s=\pi J/2$ provided 
that the condition in (\ref{gsd1}) is satisfied \cite{KR,VEWO,DM}.
In this section we choose $J=1$, so that $v_s=\pi/2$.

We will apply formula (\ref{fsc}) to the alternating XXX
spin chain in order to find the corresponding central charge
numerically. To this end we start with solving numerically 
eqs.~(\ref{be3}) in the case~$\gamma=0$. The solution of these 
equations is unique since they are extremum conditions for 
a convex function (Yang's functional, see \cite{kib} for 
the analogous consideration in the homogeneous case). 

It is interesting to notice that, if we assume that
roots of the Bethe equations form exact strings even
for finite $N$ and that the centers of these strings
are given by (\ref{be3}), then formula (\ref{fsc}) yields
the same value $\tilde{c}=2+O(N^{-1})$ for all alternating 
chains (see Table~1) thus predicting the conformal anomaly 
$\tilde{c}=2$ independent of the spins (provided
that $S_1 \neq S_2$).
The interpretation of this phenomenon is that the field
theories arising as $H=0$, $T \to 0$ limit
and  $T=0$, $H \rightarrow 0$ limit of a spin chain
are, in general, different (see the discussion in~\cite{DMN2})
and a finite-size computation based on the string hypothesis
corresponds to the latter limit. The same phenomenon occurs 
for a homogeneous chain; in this case an analytic or numeric
finite-size computation using the string hypothesis gives 
$\tilde{c}=1$ independent of spin $S$, whereas those not 
using the string hypothesis give $c=3S/(S+1)$. Comparing
this fact with our result described above, we may 
conjecture that a finite-size computation based on the 
string hypothesis performed for a chain containing 
equal number of $l$ different spins will give 
$\tilde{c}=l$ (which is the smallest possible value of $c$ 
for such a chain, see Section 5).

\begin{center}
{\small
\begin{tabular}{|c||c|c|c||c|c|}
\hline
\vphantom{ \rule{0pt}{12pt} } $S_1$,\ $S_2$ & $\lambda_\alpha^{S_1}$
 & $\lambda_\alpha^{S_2}$ & $\tilde{c}$ & $\lambda_\alpha^n$ & $c$
  \\[2pt] \hline
\vphantom{\rule{0pt}{12pt}} $\frac 12$,\ 1 &
\begin{tabular}{l}
 $\pm 0.129530$ \\
 $\pm 0.538739$
\end{tabular}
 & \begin{tabular}{l}
  $\pm 0.131152$ \\
  $\pm 0.557922$
\end{tabular}
 & 2.08 &
\begin{tabular}{l}
 $\pm 0.543466$,\  $\pm 0.129918$, \\
 $\pm 0.547454 \pm {\mathrm i}\, 0.504109$, \\
 $\pm 0.130894 \pm {\mathrm i}\, 0.500000$
\end{tabular}
 & 2.06  \\ [2pt] \hline
\vphantom{\rule{0pt}{12pt}} 1,\ $\frac 32$ &
 \begin{tabular}{l}
 $\pm 0.130724$ \\
 $\pm 0.554788$
\end{tabular}
 & \begin{tabular}{l}
  $\pm 0.131252$ \\
  $\pm 0.561180$
\end{tabular}
 & 2.10 &
\begin{tabular}{l}
 $\pm 0.549334 \pm {\mathrm i}\, 0.552225$,\\
 $\pm 0.130456 \pm {\mathrm i}\, 0.516086$,\\
 $\pm 0.562630$,\
 $\pm 0.561099 \pm {\mathrm i}\, 1.007705$,\\
 $\pm 0.131684$,\ $\pm0.131684 \pm {\mathrm i}\, 1.000000$
\end{tabular}
 & 2.59 \\ [2pt] \hline
\vphantom{\rule{0pt}{12pt}} $\frac 12$,\ $\frac 32$ &
 \begin{tabular}{l}
 $\pm 0.130351$ \\
 $\pm 0.535032$
\end{tabular}
 & \begin{tabular}{l}
  $\pm 0.131852$ \\
  $\pm 0.552153$
\end{tabular}
 & 2.08 &
\begin{tabular}{l}
 $\pm 0.449421$,\ $\pm 0.094976$,\\
 $\pm 0.668491$,\ $\pm 0.544016 \pm {\mathrm i}\, 1.055667$, \\
 $\pm 0.168193$,\ $\pm 0.131555 \pm {\mathrm i}\, 1.017436$
\end{tabular}
 &  2.60   \\ [2pt] \hline
\end{tabular}
}
\end{center}

\vspace*{2mm} \noindent {\small {\ Table 1. }
Alternating $XXX$ chains with $N=16$ sites.
$\lambda_\alpha^{S_i}$ are the real roots of eq.~(\ref{be3})
for $\gamma \to 0$; the estimate $\tilde{c}$ of the central
charge is computed by (\ref{fsc}) under the assumption that 
$\lambda_\alpha^{S_i}$ are centers of exact strings.
$\lambda_\alpha^n$~are the complex roots of eq.~(\ref{BEz})
for $\gamma \to 0$ corresponding to the genuine ground 
state; $c$ is the corresponding finite-size estimate 
computed by~(\ref{fsc}).}

Although, as we discussed above, the string solutions 
constructed on the base of (\ref{be3}) are not exact,
they, as seen from Table~1, provide good initial data 
for iterative numerical computations. Using them as a 
first approximation and applying a Newton-type method,
we solved numerically the XXX version of eqs.~(\ref{BEz})
for several alternating chains and computed the corresponding 
ground state energy. Then we used formula~(\ref{fsc}) and 
eq.~(\ref{gse1z}) in order to find the finite-size estimate of 
the central charge for these chains. Our results, presented 
in Table~2, are described by the following expression
\begin{equation}\label{css}
  c = \frac{3S_1}{S_1+1} + \frac{3(S_2-S_1)}{S_2-S_1+1} \,.
\end{equation}
This formula was conjectured in \cite{AM} as an
extrapolation of the results for the $(\frac 12, S)$
XXX chain. Since it is known that for the homogeneous
XXZ chain the central charge does not depend on the
anisotropy (if $2\gamma S < \pi$) and coincides with that of 
the XXX chain, one can expect that (\ref{css}) applies to the 
alternating XXZ chain as well (provided that the condition 
in~(\ref{gsd1}) is satisfied). This is indeed so, 
as we will demonstrate below.

\begin{center}
{\small
\begin{tabular}{|c||c|c|c|c|c|c|c|c|}
\hline
 \vphantom{ \rule{0pt}{12pt} }
 & $N=8$ \span\omit 
 & $N=16$ \span\omit  & $N=24$ \span\omit &
 $N=\infty$ \span\omit \\ [2pt] \hline
\vphantom{ \rule{0pt}{12pt} } $S_1$,\ $S_2$ & $|e|$ & $c$ &
 $|e|$ & $c$ & $|e|$ & $c$ & $|e|$ & $c$ \\[2pt] \hline
\vphantom{\rule{0pt}{12pt}} $\frac 12$,\ 1 & 
 0.666301 & 2.21 & 0.644511 & 2.06
 & 0.640791 & 2.03  & 0.637888 & 2 \\ [2pt] \hline
\vphantom{\rule{0pt}{12pt}} 1,\ $\frac 32$ &  
 0.917421 & 2.79 & 0.889926 & 2.59
 & 0.885257 & 2.55 & 0.881620 & 2.5 \\ [2pt] \hline
\vphantom{\rule{0pt}{12pt}} $\frac 32$,\ 2 & 
 1.086631 & 3.14 & 1.055557 & 2.91
 & 1.050305 & 2.86  & 1.046222 & 2.8 \\ [2pt] \hline
\vphantom{\rule{0pt}{12pt}} 2,\ $\frac 52$ & 
 1.213910 & 3.39 & 1.180345 & 3.12 &
 1.174695 & 3.07 & 1.170311 & 3 \\[2pt] \hline
\vphantom{\rule{0pt}{12pt}} $\frac 12$,\  $\frac 32$ &
 0.661489 & 2.84 &
 0.633361 & 2.60 & 0.628649 & 2.56 & 0.625 & 2.5  \\[2pt] \hline
\vphantom{\rule{0pt}{12pt}} $\frac 12$,\  2 &
 0.666989 & 3.24 &
 0.634793 & 2.94 & * & & 0.625352 & 2.8  \\[2pt] \hline
\vphantom{\rule{0pt}{12pt}} $\frac 12$,\  $\frac 52$ &
 0.675916 & 3.52 &
 0.640833 & 3.17 & * & & 0.630647 & 3  \\[2pt] \hline
\end{tabular}
}
\end{center}
\vspace*{2mm} 

\noindent {\small {\ Table 2. }
 Alternating $XXX$ chains with $N$ sites (the coupling
 constant is $J=1$). $e$ is the energy per site for the
 numerically found ground state solution of eq.~(\ref{BEz})
 for $\gamma \to 0$; $c$ is the corresponding finite-size 
 estimate  of the central charge  obtained by eq.~(\ref{fsc}).  
 The $N=\infty$ values are predicted by 
 eqs.~(\ref{gse1z}) and (\ref{css}). 
 The asterisks indicate lack of stable numerical results.
}

\section{Thermodynamics}
In this section we will investigate the thermodynamics of the
alternating XXZ chain in the case where the anisotropy 
parameter $\nu = {\pi \over \gamma} \geq 3$ is an integer 
number and $\nu > 2 S_i$.
Our analysis follows, with proper modifications,  the 
standard Bethe ansatz method \cite{FT2,BT,KR,VEWO,Fad1}.

\subsection{Bethe equations at $T > 0$}

Denote $\q_{i} \equiv 2S_{i}$; without loss of generality
we will always assume that $\q_2 \geq \q_1$.  Following 
\cite{VEWO,bado}, it is convenient to introduce the functions
\begin{equation}
 e_{n}(\lambda; \nu) \equiv 
 \frac{\sinh \gamma(\lambda+{{\mathrm i}n\over 2})}%
  {\sinh\gamma(\lambda-{{\mathrm i}n \over 2})} \,, \qquad
 g_{n}(\lambda;\nu) \equiv
   e_{n}(\lambda \pm {{\mathrm i}\pi\over 2 \gamma}) =
 {\cosh \gamma(\lambda +{{\mathrm i}n\over 2}) \over 
 \cosh \gamma(\lambda -{{\mathrm i}n\over 2})} \,.
\end{equation}
Then the Bethe equations (\ref{BEz}) acquire the form
\begin{equation}
 e_{\q_{1}}(\lambda_{\alpha})^{N/2}
 e_{\q_{2}}(\lambda_{\alpha})^{N/2}=
 -\prod_{\beta=1}^{M} e_2(\lambda_{\alpha}-\lambda_{\beta}) \,.
\label{BE}
\end{equation}
The spin, energy and momentum of a state are given in terms of 
the Bethe ansatz roots {$\lambda_{\alpha}$} by eqs.~(\ref{spin}) 
and~(\ref{PEz}). In this section we set $J=1/\pi$.

In the thermodynamic limit $N,M \to \infty$ the string 
hypothesis holds \cite{gt1}, and the Bethe equations become 
\be
\prod _{j= 1}^{2}X_{n\q_{j}}(\lambda_{\alpha}^n)^{N/2}&=& 
 (-)^{n}\prod_{m=1}^{\nu -1} \prod_{\beta=1}^{M_{m}}
E_{nm}(\lambda_{\alpha}^{n}-\lambda_{\beta}^{m})
\prod_{\beta=1}^{M_{0}}G_{n1}(\lambda_{\alpha}^{n} -
  \lambda_{\beta}^{0}) \,, \label{K1a} \\
\prod _{j= 1}^{2}g_{\q_{j}}(\lambda_{\alpha}^0)^{N/2}&=& 
 -\prod_{m=1}^{\nu -1} \prod_{\beta=1}^{M_{m}}
G_{1m}(\lambda_{\alpha}^{0}-\lambda_{\beta}^{m})
\prod_{\beta=1}^{M_{0}}e_{2}(\lambda_{\alpha}^{0}
 -\lambda_{\beta}^{0}) \,,
\label{K1b} 
\ee 
where the notation for the centers of strings are as in 
(\ref{STR}), $n$ takes values $1,2,\ldots,(\nu -1)$, and
\begin{eqnarray}
 X_{nm}(\lambda) &=& e_{|n-m+1|}(\lambda) \,
 e_{|n-m+3|}(\lambda)\ldots e_{(n+m-3)}(\lambda) \, 
 e_{(n+m-1)}(\lambda) \,, \nonumber\\
 E_{nm}(\lambda) &=& e_{|n-m|}(\lambda) \, 
 e_{|n-m+2|}^{2}(\lambda) \ldots e_{(n+m-2)}^{2}(\lambda) \,
 e_{(n+m)}(\lambda) \,, \nonumber\\
 G_{nm}(\lambda) &=& g_{|n-m|}(\lambda) \, 
 g_{|n-m+2|}^{2}(\lambda) \ldots g_{(n+m-2)}^{2}(\lambda) \,
 g_{(n+m)}(\lambda) \,.
\end{eqnarray}
The energy (\ref{PEz}) of a given state takes the form
\begin{equation}
 E = - {1\over 2}\sum_{n=1}^{\nu-1}\sum_{\alpha=1}^{M_{n}}
 (Z_{n\q_{1}}(\lambda_{\alpha}^{n})+
 Z_{n\q_{2}}(\lambda_{\alpha}^n))- {1\over 2}\sum_{\alpha=1}^{M_{0}}
 (b_{\q_{1}}(\lambda_{\alpha}^{0})+
 b_{\q_{2}}(\lambda_{\alpha}^0))-HS_{z} \,,
\label{energy}
\end{equation}
where 
\be
 Z_{nm} (\lambda) = \frac{\rm i}{2\pi} \frac{d}{d\lambda}
 \ln X_{nm} (\lambda) \,, \qquad
 b_n (\lambda) = \frac{\rm i}{2\pi} \frac{d}{d\lambda} 
 \ln g_n (\lambda) \,.
\ee
It is useful to notice that the Fourier transforms of these
quantities are 
\begin{eqnarray}
 && \hat Z_{nm}(\omega)= { \sinh \Bigl (( \nu - \max(n,m))
 {\omega \over 2} \Bigr ) \sinh \Bigl ((\min(n,m))
 {\omega \over 2} \Bigr ) \over 
 \sinh ({\nu \omega \over 2}) \sinh({\omega \over 2})} \,, 
 \label{Z} \\
 && \hat b_{n}(\omega; \nu) = -{\sinh ({n\omega \over 2}) \over 
 \sinh ({\nu \omega \over 2})} , ~~0<n< \nu,\quad 
 = -{\sinh \Big ( (n-2 \nu){\omega \over 2} \Bigr ) 
 \over \sinh ({\nu \omega \over 2})} , ~~\nu<n<3\nu \,. 
 \label{b}
\end{eqnarray}
In the $N \rightarrow \infty$ limit we obtain from
(\ref{energy}) an expression for the energy density:
\be 
 e &\equiv& \frac{E}{N} =
 -{1 \over 2} \sum_{n=1}^{\nu -1} 
 \int_{-\infty}^{\infty}d\lambda
 (Z_{n\q_{1}}(\lambda) +Z_{n\q_{2}}(\lambda))\rho_{n}(\lambda)
 -{1\over 2} \int_{-\infty}^{\infty}d\lambda (b_{\q_{1}}(\lambda)
 +b_{\q_{2}}(\lambda))\rho_{0}(\lambda) \non\\ 
 &&  + H \, \Bigl(
 \sum_{n=1}^{\nu -1} n\int_{-\infty}^{\infty}d\lambda
 \rho_{n}(\lambda) +  \int_{-\infty}^{\infty}d\lambda
 \rho_{0}(\lambda) - {1\over 4} (\q_{1}+\q_{2}) \Bigr) \,,
\label{energy2}  
\ee 
where $\rho_{n}$ is the density of the $n$-strings 
(pseudo--particles) and $\rho_{0}$ is the density of the negative 
parity string. The energy density of the ground state 
($T=0$, $H =0$), which consists of two filled Dirac seas 
(strings of length $\q_{1}$ and $\q_{2}$) is given by 
\be 
 e_{0} ={E_{0} \over N}=- {1\over 4}\sum_{i,j=1}^{2}
 \int_{-\infty}^{\infty}d\lambda \,
 Z_{\q_{i}\q_{j}}(\lambda) \, s(\lambda) \,, 
\label{energygr} 
\ee
where (cf.~(\ref{gsd1})) 
\be 
s(\lambda) ={1\over 2 \cosh(\pi \lambda)} \,, \qquad
 \hat s(\omega) = {1\over 2 \cosh{\omega \over 2}} \, . 
\ee
Besides the densities of pseudo--particles we need to
introduce the densities of holes $\tilde \rho_{n}$ (for
$n$-strings) and~$\tilde \rho_{0}$ (for the negative
parity string). Then, taking the logarithm of 
(\ref{K1a})-(\ref{K1b}) and differentiating, one finds 
\begin{eqnarray}
 && \tilde \rho_{n}(\lambda) = {1\over 2}( Z_{n\q_{1}}(\lambda)
 + Z_{n\q_{2}}(\lambda)) - \sum_{m=1}^{\nu -1}A_{nm}
 * \rho_{m}(\lambda) - B_{n} * \rho_{0}(\lambda) \,, \non \\
 && -(\rho_{0}(\lambda) + \tilde \rho_{0}(\lambda)) = 
 {1\over 2}(b_{\q_{1}} +b_{\q_{2}}) -\sum_{m=1}^{\nu -1}B_{m} 
 * \rho_{m}(\lambda) -a_{2}*\rho_{0}(\lambda) \,, 
\label{densities} 
\end {eqnarray}
where
\begin{eqnarray}
 && A_{nm} (\lambda) = \delta_{nm} \delta(\lambda) +
 \frac{\rm i}{2\pi} \frac{d}{d\lambda}
 \ln E_{nm} (\lambda) \,, \quad
 B_{n} (\lambda) = \frac{\rm i}{2\pi} \frac{d}{d\lambda} 
 \ln G_{1n} (\lambda) \,, \quad
 a_n (\lambda) = \frac{\rm i}{2\pi} \frac{d}{d\lambda} 
 \ln e_n (\lambda) \,,  \non \\ [0.5mm]
 && \hat A_{nm}(\omega)=
 2 \cosh \bigl(\frac{\omega}{2}\bigr) \, 
       \hat{Z}_{nm}(\omega) \,, \qquad
\hat B_{n}(\omega)=-{2\cosh ({\omega \over 2})
 \sinh \bigl ({n\omega \over 2}\bigr)
\over \sinh ({\nu \omega \over 2})} + \delta_{n,\nu-1}\,. 
\label{fourier}
\end{eqnarray}
In order to find $\rho_{n}$, $\rho_{0}$ in terms of 
$\tilde \rho_{n}$, $\tilde \rho_{0}$, it is convenient to 
consider the convolution of the first of the density 
equations (\ref{densities}) with the inverse of $A_{nm}$
which is given by 
\be \label{Ainv}
 \hat A_{nm}^{-1} = \delta_{nm} -\hat s(\omega)(\delta_{n,m+1} 
 +\delta_{n,m-1}) \,.
\ee
Taking into account the following identities (summation
over $m$ is implied)
\be 
\label{azb}
& A_{nm}^{-1} * Z_{mk}(\lambda) =
s(\lambda) \, \delta_{nk} \,, \quad 
 A_{nm}^{-1} * m  = {\nu \over 2} \delta_{n,\nu -1} \,,\quad
 A_{nm}^{-1} * B_{m}(\lambda) = - s(\lambda) \, 
 \delta_{n,\nu -2} \,, & \\ [1.5mm]
&  a_2 (\lambda) + s * B_{\nu-2} (\lambda) = 0 \,, \qquad
 b_n (\lambda) + s(\lambda) \, \delta_{n,\nu-1} = 
 - s * Z_{n,\nu-2} (\lambda) = s * B_n(\lambda) \,, &
\label{azb2}
\ee  
we obtain the following expressions
\be
\rho_{n}(\lambda) &=& {1 \over 2} s(\lambda) (\delta_{n\q_{1}}
 +\delta_{n\q_{2}}) - \sum_{m=1}^{\nu -1} A_{nm}^{-1}* 
 \tilde \rho_{m}(\lambda)  
 +\delta_{n, \nu -2} s * \rho_{0}(\lambda)
 \,, \non\\
\rho_{0}(\lambda) &=&- \tilde \rho_{0}(\lambda) 
 +s*\tilde \rho_{\nu -2}(\lambda) +
 {1 \over 2} s(\lambda) (\delta_{\q_{1},\nu-1}
 +\delta_{\q_{2},\nu-1}) \,.
\label{densities2} 
\ee
According to the standard statistical principles, 
the entropy density of the system is given by 
\be 
 s \equiv {S \over N} = \sum_{n=0}^{\nu -1}
\int_{-\infty}^{\infty} d \lambda \Big (\rho_{n}(\lambda)
\ln(1+{\tilde \rho_{n}(\lambda) \over \rho_{n}(\lambda)}) + 
\tilde\rho_{n}(\lambda) \ln(1+{\rho_{n}(\lambda) \over 
\tilde \rho_{n}(\lambda)})\Big ) \,.
\label{entropy1}
\ee
Denote $\eta_{n}(\lambda) ={ \tilde \rho_{n}(\lambda) 
 \over \rho_{n}(\lambda)}$. The free energy of the system
is $F=E-TS$.
The equilibrium condition, $\delta F =0$, along with relations 
(\ref{energy2}), (\ref{densities}), and (\ref{entropy1})
yields the thermodynamic Bethe ansatz equations
\be 
 T \ln \Big (1+\eta_{n}(\lambda)\Big ) &=& nH 
 -{1\over 2} (Z_{n\q_{1}}(\lambda)+Z_{n\q_{2}}(\lambda))  \non\\
 && + T\sum_{m=1}^{\nu -1} A_{nm}* 
   \ln \Big (1+\eta_{m}^{-1}(\lambda)\Big ) 
 - T B_{n}*\ln \Big (1+\eta_{0}^{-1}(\lambda)\Big ), \non\\ 
 T \ln \Big ({1+\eta_{0}(\lambda) 
   \over 1+\eta_{0}^{-1}(\lambda)}\Big ) &=& H
 -{1 \over 2}(b_{\q_{1}}(\lambda)+b_{\q_{2}}(\lambda)) \non\\
 && + T \sum_{m=1}^{\nu -1} B_{m}* 
 \ln \Big (1+\eta_{m}^{-1}(\lambda)\Big)  
 - T a_{2}*\ln \Big (1+\eta_{0}^{-1}(\lambda)\Big ).
\label{TBA} 
\ee
These equations can be rewritten with the help of
(\ref{Ainv})--(\ref{azb2}) as follows
\be 
\epsilon_{n}(\lambda) &=&  s(\lambda)*T 
 \ln [(1+ \eta_{n+1}(\lambda))(1+\eta_{n-1}(\lambda))]
 + {\nu H \over 2} \, \delta_{n, \nu-1} \non\\ 
 && +\ \delta_{n, \nu-2} \, s(\lambda)*T 
 \ln \Big( 1+ \eta_{0}^{-1}(\lambda) \Big ) 
 -{1\over 2} s(\lambda) \,(\delta_{n\q_{1}}
 +\delta_{n\q_{2}}) \,, \label{TBA2} \\
\epsilon_{0}(\lambda) &=& {\nu H \over 2}
 - s(\lambda)*T \ln (1+ \eta_{\nu -2}(\lambda)) 
 + {1\over 2} s(\lambda) \,(\delta_{\q_{1},\nu-1}
 +\delta_{\q_{2},\nu-1}) \,, \non
\ee
where $\epsilon_{n}(\lambda) \equiv T \ln \eta_{n}(\lambda) $
are the so-called pseudo-energies (the terms with 
$\ln (1+\eta_0)$ and $\ln (1+\eta_\nu)$ are omitted in the 
equations for $\epsilon_{1}$ and $\epsilon_{\nu-1}$, respectively). 
Observe that the equations for $\epsilon_{0}$ and $\epsilon_{\nu-1}$ 
in (\ref{TBA2}) lead to the relation
\be
  \epsilon_{0}(\lambda) = \nu H - \epsilon_{\nu-1} (\lambda) \,.
\label{ee} 
\ee
This implies, in particular, that the case $\q_2=\nu{-}1$ (recall 
that $\q_1 \leq \q_2$) is special because then an extra term appears 
in the equation for~$\epsilon_{0}$. Indeed, if $\q_2 = \nu{-}1$, 
the $\q_2$-strings get close to the odd parity strings, namely
$\Im ( \lambda_{\beta}^{(0)} -\lambda_{\alpha}^{(\q_2,1)}) = 1$ 
(see~(\ref{STR})). As seen from the r.h.s.~of (\ref{BEz}), this 
affects the structure of the roots of the Bethe equations.

Using (\ref{densities}), (\ref{TBA}), and (\ref{azb2}),
we obtain the following expression for the free energy density
at the equilibrium
\begin{eqnarray} 
 f &=& {F \over N} = 
 - {T \over 2}\sum_{n=1}^{\nu-1} \int_{-\infty}^{\infty} 
 d \lambda \ln(1+\eta_{n}^{-1}(\lambda))(Z_{n\q_{1}}(\lambda)
 +Z_{n\q_{2}}(\lambda)) \non\\ 
 && +\ {T \over 2}\int_{-\infty}^{\infty}
 d \lambda \ln(1+\eta_{0}^{-1}(\lambda))(b_{\q_{1}}(\lambda)
 +b_{\q_{2}}(\lambda)) - {(\q_{1} +\q_{2}) H \over 4} 
 \non \\ [0.5mm]  
 &=& e_0 - {T\over 2} \sum_{i=1}^2 \Bigl\{
 \int_{-\infty}^{\infty} d \lambda \, s(\lambda)
 \ln(1+\eta_{\q_{i}}(\lambda))  +
 \delta_{\q_i,\nu-1}  \int_{-\infty}^{\infty} d \lambda \,
 s(\lambda)\ln(1+\eta^{-1}_{0}(\lambda)) \Bigr\} \,.
\label{fe}
\end{eqnarray}

\subsection{The $T=0$, $H \ll 1$ limit}
At zero temperature, the ground state in the anti ferromagnetic 
regime consists of two filled Dirac seas (strings of length 
$\q_{1}$ and~$\q_{2} \geq \q_{1}$). In this case we can rewrite 
the equations (\ref{TBA}) for the strings $\q_{1}$
and $\q_{2}$ only in the following approximate form 
\begin{eqnarray}
 && \epsilon_{i}(\lambda) = -{1 \over 2}\sum_{j=1}^{2}
 Z_{\q_{i}\q_{j}}(\lambda) + \q_{i} H +
 \sum_{j=1}^{2}\tilde A_{\q_{i}\q_{j}}* T
 \ln(1+\eta_{\q_{j}}^{-1}(\lambda)) \,, 
\label{appr} \\ 
 && \tilde A_{nm}(\lambda) \equiv A_{nm}(\lambda)
  -\delta_{nm} \delta(\lambda) \,. \non  
\end{eqnarray}
In the $T=0$ limit the following quantities are defined
for $i=1,2$
\be 
 T \ln(1+\eta_{\q_{i}}^{\pm 1}) \to \pm \epsilon_{i}^{\pm} \,, 
 \qquad 
 \epsilon_{i}^{\pm} = {1\over 2}(\epsilon_{i} \pm|\epsilon_{i}|) \,.
\ee 
Then equation (\ref{appr}) can be rewritten as follows 
\be 
 \sum_{j=1}^{2}A_{\q_{i}\q_{j}} *\epsilon_{j}(\lambda) = 
 -{1 \over 2}\sum_{j=1}^{2} Z_{\q_{i}\q_{j}}(\lambda) + \q_{i} H +
 \sum_{j=1}^{2}\tilde A_{\q_{i}\q_{j}}* \epsilon_{j}^{+}(\lambda) \,.
\label{appr2} 
\ee 
By solving the above system with the help of (\ref{azb}),
we obtain  
\be \label{appr3}
\epsilon_{i}(\lambda) = -{1 \over 2}s(\lambda) +
 \frac 12  \delta_{i2} \, \xi H + 
 M_{ij}* \epsilon_{j}^{+}(\lambda) \,,
 \qquad \xi \equiv {\nu \over \nu -\q_{2}} \,, 
\ee 
where the kernel $M$ has the following Fourier image
\be \label{matrix1}
\hat M(\omega)=\left(
           \begin{array}{cc}
            \hat h_{1}(\omega) &\hat h(\omega)     \\
            \hat h(\omega)     &\hat h_{2}(\omega)\\
                               \end{array}\right) \,,
\ee
with
\be 
\hat h_{1}(\omega) &=& {\sinh (\delta \q -1){\omega \over 2}
\over 2\cosh{\omega \over 2}\sinh\delta \q{\omega \over 2}} 
+{\sinh (\q_{1}-1){\omega \over 2} \over 2 \cosh{\omega \over
2}\sinh(\q_{1}){\omega \over 2}}, \non\\ \hat h_{2}(\omega)
&=&{\sinh (\delta \q -1){\omega \over 2} \over 2\cosh{\omega \over
2}\sinh \delta \q{\omega \over 2}}+ {\sinh (\nu - \q_{2}-1){\omega \over
2} \over 2 \cosh{\omega \over 2}\sinh (\nu -\q_{2}){\omega \over
2}},~~\hat h(\omega) = {\sinh {\omega \over 2} \over 2\cosh{\omega
\over 2}\sinh \delta \q{\omega \over 2}},
\ee 
and $\delta \q = \q_{2} -\q_{1}$. Notice that, as in \cite{doikou}, 
the quantities $h_{i}$, $h$ are related to the LL (RR) scattering
between left (right) movers, and the LR (left--right movers) 
massless scattering (see e.g.~\cite{zz}).

Eqs.~(\ref{appr3}) imply that $\epsilon_{i}(\lambda)$ are
monotonically increasing functions of~$|\lambda|$ with the
maximal values given by
\be
 \epsilon_{i}(\infty) = {1 \over 2} \xi H \,
 \Bigl( \bigl( 1- \hat{M}(0) \bigr)^{-1} {0 \choose 1} \Bigr)_i 
 = \q_i H\,. 
\label{einf}
\ee
Therefore, for $H$ small but non-vanishing, the pseudo-energies 
$\epsilon_{i}(\lambda)$ are negative only on finite intervals
$|\lambda|< \alpha_i$. Having (\ref{appr3}) in mind, we introduce
the zeroes of the pseudo-energies as follows 
\be 
 \alpha_{i} = -{1 \over \pi} (\ln \xi H + \ln k_{i} +O(1/\ln H)) \,,
\label{zeroes} 
\ee 
where $k_i$ (which do not depend on $H$) are to be determined later. 
Next we define the functions 
\be
 S_{i}(\lambda) = (\xi H)^{-1} \epsilon_{i}(\lambda + \alpha_{i})
 \qquad \mathrm{for}\ \lambda \geq 0
\ee 
and $ S_{i}(\lambda)=0$ for $\lambda <0$. Making the shift
$\lambda \to \lambda + \alpha_i$ in (\ref{appr3}) and using that 
$s(\lambda + \alpha_i) \to \xi H \, k_i \, e^{-\pi\lambda}$
(since $\alpha_{i} \to \infty$ as $H \to 0$), we obtain
a Wiener--Hopf type system
\be 
S_{1}(\lambda)&=&- {k_{1}\over 2}e^{-\pi \lambda} 
+ \int_{0}^{\infty}d \lambda' h_{1}(\lambda -\lambda') 
S_{1}(\lambda') + \int_{0}^{\infty}d \lambda' h(\lambda 
-\lambda'+\alpha_{1}-\alpha_{2}) S_{2}(\lambda') ,
\non\\S_{2}(\lambda)&=& {1\over 2} -{k_{2} \over 2}e^{-\pi \lambda} +
\int_{0}^{\infty}d \lambda' h(\lambda
-\lambda'+\alpha_{2}-\alpha_{1})S_{1}(\lambda') + 
\int_{0}^{\infty}d \lambda'
 h_{2}(\lambda -\lambda')S_{2}(\lambda'). 
\label{system} 
\ee 
The above system can be solved via the standard Wiener--Hopf 
methods as follows. The Fourier image of the matrix kernel $K$
of the system (\ref{system}) is
\be 
 \hat K(\omega)=\left( \begin{array}{cc}
  \hat h_{1}(\omega) &
   e^{-{\mathrm i}\omega(\alpha_{1} -\alpha_{2})}\hat h(\omega)  \\
  e^{-{\mathrm i}\omega(\alpha_{2} -\alpha_{1})}\hat h(\omega)     
    &\hat h_{2}(\omega)\\
          \end{array}\right)\,
          \label{matrix} .
\ee
Its resolvent admits the following factorization \cite{Krein}
\be \label{KG}
 (1-\hat K(\omega))^{-1} =G_{+}(\omega)G_{-}(\omega) \,, 
\ee
where $G_{\pm}(\omega)$ are analytic for $\pm\Im \omega \geq0$,
respectively, and also $G^{\,t}_{+}(\omega) = G_{-}(-\omega)$
for $\Im \omega \geq 0$. The solution
of the system (\ref{system}) (see also \cite{DMN1}) is 
\begin{eqnarray} 
 \hat S(\omega) &=& {{\mathrm i}\over 2}({1 \over \omega + {\mathrm i}0} 
 -{1 \over \omega +{\mathrm i}\pi}) \, G_{+}(\omega)G_{-}(0){0 \choose 1} 
\label{solution} \,, \\
 k &=& G_{-}^{-1}(-{\mathrm i}\pi)G_{-}(0){0 \choose 1} \,. 
\label{2}
\end{eqnarray}
Denote $D= \det (1-\hat K(0))$. {}From 
(\ref{matrix})--(\ref{2}) we compute the following quantity
\be 
 k^{t} \hat S({\mathrm i}\pi)= 
 {1 \over 4 \pi}\Big (0 ~1\Big ) G_{+}(0)G_{-}(0){0 \choose 1} 
 =  {1\over 4\pi}D^{-1} \bigl(1-\hat h_{1}(0) \bigr)
 = \frac{ \q_2}{2\pi \, \xi} \,.
\label{S} 
\ee 

Recall that the odd parity string contributes to the
free energy (\ref{fe}) only if $\q_i =\nu{-}1$. Observe,
however, that its contribution vanishes as $T \to 0$ because 
it follows from (\ref{einf}) and (\ref{ee}) that
$\epsilon_0(\lambda)$ is positive. Thus, as $T \to 0$, 
the free energy of the system becomes 
\begin{eqnarray} 
 f &=& e_{0} - \sum_{i=1}^{2} \int_{\alpha_{i}}^{\infty}
 d \lambda \, s(\lambda) \epsilon_{i}(\lambda) =
 e_{0} - \xi H \sum_{i=1}^{2} \int_{0}^{\infty} d \lambda 
 s(\lambda+\alpha_{i})S_{i}(\lambda) \non \\
 &=& e_{0} - \xi^2 H^2 \sum_{i=1}^{2} k_{i}
 \int_{0}^{\infty} \, d\lambda e^{-\pi \lambda} S_{i}(\lambda)
 = e_{0} - \xi^{2} H^{2}\, k^{t} \hat S({\mathrm i}\pi) \,. 
\label{111}
\end{eqnarray}
Substituting here (\ref{S}), we obtain the following expression 
for the magnetic field dependence of the free energy, 
\be 
 f= e_{0} - {1 \over 2 \pi}{\nu \q_{2} \over \nu -\q_{2}}H^{2} \,.
 \label{free1}
\ee

The magnetic susceptibility is given by $\chi = -\partial^2_H f$.
The value of $v_s \chi$ does not depend on the normalization
of the Hamiltonian ($v_s$ is the speed of sound, which in this
section is $\frac 12$). Thus, we infer from (\ref{free1}) that 
\be\label{chi}
 v_s \, \chi = {1\over 2\pi}{\nu \q_{2} \over \nu -\q_{2}} \,. 
\ee
Notice that the contribution of the external magnetic field to 
the free energy (\ref{free1}) does not depend on $\delta \q$, 
that is why our result coincides with the magnetization of the 
homogeneous XXZ chain found in~\cite{BT} (where $v_s=J\pi/2$). 
Also, choosing $\q_2=2$ in (\ref{chi}), we recover the results 
obtained for the alternating ($\frac 12$,1) chain both in 
the XXX \cite{DMN1} and XXZ \cite{DM} cases.

\subsection{The $H=0$, $T \ll 1$ limit}
Here we will investigate the low temperature thermodynamics 
of the model in order to find the corresponding conformal anomaly.
In order to evaluate the entropy of the system at small but
nonzero temperature we use the following approximations 
which hold for $\lambda \gg 1$ (see, e.g.,~\cite{BT,KR}) 
\be 
 \rho_{n}(\lambda) = {1 \over \pi} (-1)^{\delta_{n0}}
 f_{n}(\lambda) {d \over d \lambda} 
   \epsilon_{n} (\lambda) \,, \qquad 
 \tilde \rho_{n}(\lambda) = {1 \over \pi} (-1)^{\delta_{n0}}
 (1-f_{n}(\lambda)) {d \over d \lambda} \epsilon_{n}(\lambda) \,,
\label{def} 
\ee 
where 
\be
 f_{n}(\lambda) = (1+e^{{\epsilon_{n} \over T}})^{-1} \,.
\label{appr1} 
\ee
Employing (\ref{def}) and taking into account that 
$\epsilon_{n}(-\lambda)=\epsilon_{n}(\lambda)$,
 the entropy (\ref{entropy1}) can be written as 
\be 
 s &=& -{2 \over \pi} \sum_{n=0}^{\nu-1} (-1)^{\delta_{n0}}
\int_{\epsilon_{n}(0)}^{\epsilon_{n}(\infty)}d\epsilon_{n}
\Big (f_{n}(\lambda) \ln f_{n}(\lambda) +(1-f_{n}(\lambda) \ln
(1-f_{n}(\lambda))\Big ) \non \\
 &=& {4T \over \pi}  \sum_{n=0}^{\nu -1} (-1)^{\delta_{n0}} \Big
 (L(f_{n}^{(0)})-L(f_{n}^{(\infty)})\Big) \,,
 \label{entropy2}
\ee 
where $f_{n}^{(0)}\equiv f_n(0)$, 
$f_{n}^{(\infty)}\equiv f_n(\infty)$, and
we used the Rogers dilogarithm defined as 
\be \label{Rog}
 L(x) = -{1\over 2} \int_{0}^{x} dy \,
 \Bigl( {\ln y \over 1-y} + {\ln (1-y) \over y} \Bigr) \,.
\ee 
This function satisfies the following relations
\be\label{abel}
 L(x) + L(1-x) = L(1) \,, \qquad
 L(x^2) = 2 L(x) - 2 L(\frac{x}{1+x}) \,,
\ee
the second of them is called the Abel duplication formula.
Moreover, the following identities hold (the first is an
easy consequence of the Abel formula; for a proof of the second 
one see, e.g.,~\cite{KR})
\be 
 \sum_{k=2}^{\nu- 1}L({1\over k^{2}}) +2L({1\over \nu}) 
 = L(1) \,, \qquad
  \sum_{k=1}^{n} L\Bigl( \frac{\sin^2({\pi \over n+3})}%
 {\sin^2({\pi (k+1) \over n+3})}\Bigr) = 
 \frac{2 n}{n+3} \, L(1) \,,
\label{i1} 
\ee 
where $n\geq 1$ and $\nu \geq 2$ are integer.

The knowledge of the entropy allows one to find
the corresponding specific heat,
$C_{u} = T \partial_T s(T)$. 
On the other hand, at low temperature, we have \cite{bc}, 
$ C_{u} = {\pi c \over 3 v_s }T + o(T)$,
where $c$ is the effective central charge of the conformal 
field theory related to the $T=0$ limit, and $v_s$ is the 
speed of sound (Fermi velocity), which is $v_s= {1 \over 2}$ 
for our choice of the coupling constant $J=1/\pi$.
Thus, taking into account that $L(1) = \pi^2/6$, we
infer from (\ref{entropy2}) that for the alternating
chain we have
\be
 c(\nu | \q_1,\q_2) = {1 \over L(1)}  \sum_{k=1}^{\nu} \Big
 (L(f_{k}^{(0)})-L(f_{k}^{(\infty)})\Big) \,,
 \label{ceff}
\ee
where we introduced a new variable,
$f_\nu(\lambda) \equiv 1-f_0 (\lambda)$,
and used the first property in~(\ref{abel}). 
Observe that the replacement of the quasi-particle 
corresponding to the negative parity by a formal $k=\nu$ 
particle makes eqs.~(\ref{TBA2}) to look uniformly 
(since $\epsilon_\nu = -\epsilon_0$, $\eta_\nu = 1/\eta_0$).
Furthermore, the formal \hbox{$\nu$-th} particle contributes to 
the central charge (\ref{ceff}) in the same way as a genuine 
quasi-particle. This allows us, in the following analysis, to 
treat all the $k=1,\ldots,\nu$ particles on the equal footing.

For $T \ll 1$, the functions $\ln (1+\eta_n(\lambda))$ 
vary very slowly for $|\pi\lambda| \ll - \ln T$ and
for $|\pi\lambda| \gg - \ln T$. Therefore, at $H=0$,
$T \ll 1$, and $\lambda \to \infty$, eqs.~(\ref{TBA2})
turn into the following system (which is the same as in 
the homogeneous XXZ case~\cite{BT})
\be
 f_k^{-1} -1 = 
 \prod_{m=1}^\nu  \bigl( f_m \bigr)^{-\frac 12 I_{k m}}
 \,, \qquad k=1,\ldots,\nu\,,
\label{differ}
\ee
where $f_k=f_k^{(\infty)}$, and $I_{km}$ is the incidence 
(adjacency) matrix of the graph (that coincides with the 
Dynkin diagram of $D_n$ with $n=\nu$)
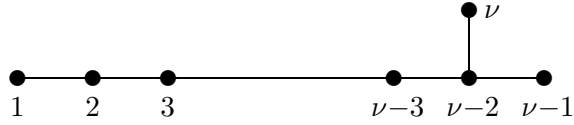
\begin{center}
\setlength{\unitlength}{1mm}
\begin{picture}(80,20)
 \put(0,10){\line(1,0){70}}
 \put(0,10){\circle*{2}}
 \put(10,10){\circle*{2}}
 \put(20,10){\circle*{2}}
 \put(50,10){\circle*{2}}
 \put(60,10){\circle*{2}}
 \put(70,10){\circle*{2}}
 \put(60,10){\line(0,1){9}}
 \put(60,19){\circle*{2}}
 \put(-1,5){$1$}
 \put(9,5){$2$}
 \put(19,5){$3$}
 \put(47,5){$\nu{-}3$}
 \put(57,5){$\nu{-}2$}
 \put(67,5){$\nu{-}1$}
 \put(62,18){$\nu$}
\end{picture}
\end{center}
Under the restriction $0 \leq f_n(\lambda) \leq 1$, 
the system (\ref{differ}) has a unique solution:
\be 
 f_{k}^{(\infty)} = {1\over(k+1)^{2}} \,, 
  \quad k \leq \nu-2 \,,  \qquad
 f_{\nu-1}^{(\infty)} = f_{\nu}^{(\infty)} = {1\over \nu} \,. 
\label{sol1} 
\ee
It follows from the first relation in (\ref{i1}) that
\footnote{Which means, in view of (\ref{abel}) and the
relation $f_\nu=1-f_0$, that the pseudo-energies at 
$\lambda=\infty$ do not contribute to the central charge.} 
\be
 \sum_{k=1}^{\nu} L(f_{k}^{(\infty)}) = L(1) \,.
\label{Dn}
\ee

A remark is in order here. In the thermodynamic Bethe
ansatz approach \cite{KM} to the simply laced affine 
Toda models (when the minimal part of the scattering
matrix is considered), the corresponding effective
central charge $c[{\mathfrak g}]$ is given by~(\ref{ceff}), 
where $f_k^{(\infty)}=0$ and the values of $f_k^{(0)}$ satisfy 
the system (\ref{differ}) with $I_{km}$ being the incidence 
matrix of the related Lie algebra~$\mathfrak g$.
{}From this point of view, equation (\ref{Dn}) represents
the  fact that $c[D_n]=1$, independent of~$n$. 
In the case of ${\mathfrak g} = A_n$, the system (\ref{differ}) 
is solved by \cite{BT,BR}
\be
 f_{k} &=& {\sin ^{2}({\pi \over n+3}) \over 
 \sin^{2}({\pi (k+1) \over n+3})} \,, 
 \qquad k=1, \ldots ,n \,,
\ee
which, according to the second identity in (\ref{abel}), implies
\be
  c[A_n] = \frac{2n}{n+3} \,,
 \label{An}
\ee
as is well known for the $A_n$ affine Toda models.

Now we analyze equations (\ref{TBA2}) for $T \ll 1$ and 
$\lambda \to 0$. Consider first the case when 
$\q_1 \leq \q_2 < \nu-1$.
The presence of the $s(\lambda)\delta_{n \q_i}$ 
terms in (\ref{TBA2}) leads to 
$\epsilon_{n}/T = -\infty$ for $n=\q_1,\q_2$. Thus, we have
$f^{(0)}_{\q_1}=f^{(0)}_{\q_2}=1$, which implies the 
following modification of the system~(\ref{differ})
\be
 \bigl( f^{(0)}_k \bigr)^{-1} -1 = \sigma_k \,
 \prod_{m=1}^\nu  \bigl( f^{(0)}_m \bigr)^{-\frac 12 I_{k m}}
 \,, \qquad k=1,\ldots,\nu\,,
\label{differ2}
\ee
where $\sigma_k=(1-\delta_{k\q_1})(1-\delta_{k\q_2})$.

The system (\ref{differ2}) corresponds to a graph which is 
the $D_\nu$--type diagram with the $\q_1$ and $\q_2$ nodes 
removed (notice that these nodes contribute $c_0=1$ each to 
the central charge).
If $\q_2 < \nu {-} 2$, the corresponding graph 
consists of three disconnected subgraphs -- that of $A_n$ 
with $n_1=\q_1 {-} 1$, $n_2=\q_2 {-} \q_1 {-} 1$, and 
$D_n$ with $n=\nu {-} \q_2$. 
The contribution of the latter subgraph is $L(1)$ and 
it cancels in (\ref{ceff}) against~(\ref{Dn}).
Taking into account (\ref{An}), we conclude that the 
corresponding central charge is
\be
 c(\nu | \q_1,\q_2) = c[A_{\q_1-1}] + c[A_{\delta \q -1}] + 2c_0
 + c[D_{\nu - \q_2}] - c[D_\nu] = {3\q_{1} \over \q_{1} +2} +
  {3 \delta \q \over \delta \q +2} \,,
\label{central} 
\ee
which agrees with our numerical results (\ref{css}) and with the 
conjectured formula \cite{AM} for the XXX alternating chain. 
For the special case where $(S_{1},S_{2})=(\frac 12,1)$ we 
recover the result of~\cite{DMN1}. Observe that (\ref{central}) 
remains valid also for $\delta \q =0$, 
although the derivation should be slightly modified.

If $\q_2 = \nu {-}2$, the emerging $D_n$--type subgraph 
for $n=2$ is degenerate (it consists of two disconnected nodes).
But we still have $c[D_2]=1$ because in this 
case eqs.~(\ref{differ2}) yield 
$f^{(0)}_{\nu-1}=f^{(0)}_{\nu}= \frac 12$ and the 
contribution of these nodes is $L(1)$ thanks to the
fact that $L(\frac 12)=\frac 12 L(1)$. Thus,
eq.~(\ref{central}) remains correct in this case as well.
 
For $\q_2 = \nu {-} 1$, eqs.~(\ref{TBA2}) imply that 
$\epsilon_{0}/T = +\infty$ and hence 
$f^{(0)}_{\q_1}=f^{(0)}_{\q_2}=f^{(0)}_{\nu}=1$ (so
these nodes contribute $3c_0=3$ to the central charge).
Thus, we have to consider the system (\ref{differ2}) 
with \hbox{
$\sigma_k=(1-\delta_{k\q_1})(1-\delta_{k,\nu-1})(1-\delta_{k\nu})$}.
The corresponding graph consists (for $\delta \q >0$) of two $A_n$ 
diagrams with $n_1=\q_1 {-} 1$ and 
$n_2=\nu {-} \q_1 {-} 2=\delta \q {-}1$.
Therefore, we conclude that the resulting central charge
is given by the same expression~(\ref{central}).
Again, if $\delta \q {=}0$, the derivation should be slightly 
modified but (\ref{central}) remains valid.

\subsection{Remarks on the underlying CFT}
Eq.~(\ref{central}) can
be interpreted as the sum of the central charges of two $SU(2)$ 
$WZW$ models at levels $\q_{1}$ and~$\delta \q$ which indicates
that the structure of the effective conformal field 
theory is $WZW_{\q_{1}} \otimes WZW_{\delta \q}$. Furthermore,
it was argued in \cite{AM2} for the homogeneous XXZ chain that
its underlying CFT contains composite fields formed by products
of the Gaussian and the $Z(\q)$-parafermionic fields in 
agreement with the representation 
$c(\nu|\q)=1+c[A_{\q-1}]$ (cf.~(\ref{central}) for $\delta \q=0)$. 
We may expect a similar operator content for the alternating 
chain, in particular, composite fields involving $Z(\q_1)$ and
$Z(\delta \q)$ parafermions are likely to be present.

It should be remarked here that (\ref{central}) gives
the resulting central charge but does not reflect the fact
that all $\nu$ quasi-particles in (\ref{TBA}) contribute
to its value. It is however known for the thermodynamic
Bethe ansatz treatment of models with Lie algebraic symmetries,
like the simply laced affine Toda models, that the related
Lie algebraic structure is often exhibited in the
quasi-particle content of characters of the related 
CFT~\cite{KKMM,F}.  It is therefore natural to look for
a basis in the CFT related to the (alternating) XXZ 
chain that would contain $\nu$ quasi-particles. 

It was observed in \cite{KKMM} that many characters
of various conformal models admit the following
(so-called fermionic) representation,
\be
 \chi(q) = \sum_{m_1,\ldots,m_n} q^{ {1 \over 2} 
 {\mathbf m}^t B {\mathbf m} + {\mathbf m} \cdot {\mathbf a} 
 + a_0 }  \prod_{k=1}^n \left[ { ( (1-B) {\mathbf m} + 
 {\mathbf u} )_k  \atop m_k } \right]_q 
\label{char}
\ee
where $B$ is $n{\times}n$ real matrix, $\mathbf a$ and $\mathbf u$ 
are $n$-component vectors, $a_0$ is a constant, and the $q$-binomial 
coefficients are defined for $0\leq m \leq n$ as
\be
 \left[ {n \atop m} \right]_q = \left\{ 
 \begin{array}{cl} 
  { (q)_n \over (q)_m (q)_{n-m} } 
           &\quad  {\rm if} \quad n<\infty \\[1.5mm]
  { 1 \over (q)_m } & \quad  {\rm if} \quad n=\infty 
 \end{array}
 \right. , \qquad
 (q)_m = \prod_{k=1}^{m} (1-q^k) \,.
\ee
Usually $B$ is related to the Cartan matrix $C_{\mathfrak g}$ 
of some Lie algebra $\mathfrak g$, and there are typically
certain restrictions on the summation over~$m_k$. 
Analysis of the $q\to 1^-$ asymptotics of (\ref{char}) 
allows one to find the corresponding central charge
as a combination of Rogers dilogarithms with arguments
satisfying two systems of non-linear equations~\cite{KKMM}.
It turns out that our equations (\ref{ceff}), (\ref{differ}),
and (\ref{differ2}) are exactly of this form provided that
we make the following identification
\be
 n=\nu \,, \qquad B = \frac 12 C_{D_\nu} \,, \qquad
 u_{\q_1} = u_{\q_2} = \infty \,, \qquad u_\nu= u_{\nu-1} \,,
 \label{dn}
\ee
where $C_{D_\nu}$ is the Cartan matrix of the Lie
algebra~$D_\nu$.
Furthermore, characters of the type (\ref{char}) 
based on ${\mathfrak g} = A_k$ arise in the context of
the spinon basis of $SU(2)$ WZW models related to
the homogeneous XXX spin chain~\cite{BLS}. This, together 
with the correspondence (\ref{dn}), makes it plausible that
the spinon basis related to the (alternating) XXZ spin
chain with the anisotropy $\gamma=\pi/\nu$ will 
correspond to (\ref{char}) based on~${\mathfrak g} = D_\nu$.

Let us stress here that the correspondence (\ref{dn}) 
is based solely on the data obtained by the thermodynamic
analysis of the ground state of the system. It is worth 
to mention that an additional information, in particular,
about $\mathbf a$ and $a_0$ in (\ref{char}), can be 
obtained from the finite size analysis of low lying
excitations (see, e.g., \cite{VK,AM2}) because this analysis 
allows one to find the anomalous dimensions of the fields 
present in the underlying conformal theory.   

\subsection{The $H=0$, $T \gg 1$ limit}
It is also interesting to examine the high temperature behavior
of the entropy and deduce the number of states of the model. In
the high temperature limit the pseudo energies $\epsilon_{n}$
become independent of~$\lambda$~\cite{BT}. Consequently,
neglecting the free terms in (\ref{TBA2}), we obtain the
same constant TBA equations as in (\ref{differ}). Thus,
we have (cf.~(\ref{sol1}))
\be 
 1 + \eta_{n} = (n+1)^{2-\delta_{n, \nu-1}} \,, 
  \qquad  n=1,\ldots,\nu-1 \,.
 \label{nht}
\ee 
According to (\ref{ee}) and (\ref{fe}), the free energy density
at $H=0$ in the high temperature limit becomes
\be 
 f = -{T \over 4} \sum_{n=\q_{1}, \q_{2}} 
 (1 + \delta_{n, \nu-1} ) \ln (1 + \eta_{n}) \,. 
\ee 
Hence, taking into account (\ref{nht}), the entropy is given by 
\be 
S= {N \over 2}\sum_{n=\q_{1}, \q_{2}}  \ln  (n+1) \,. 
\label{s} 
\ee 
Which implies that (recall that $\q_{i} =2S_{i}$) 
the number of states for the system is
\be
 \Big ((2S_{1}+1)(2S_{2}+1)\Big )^{{N\over 2}} \,,
\label{states}
\ee
as one could have expected.

\section{Generalization to $l$ spins}
Consider the XXZ chain containing $l$ different spins,
$S_1, \ldots, S_l$, assigned to sites with periodicity~$l$, i.e, 
the spins at the sites $n$ and $(n{+}l)$ are the same. Let us 
outline how our analysis of the alternating chain extends 
to this case.

Introduce the following set of transfer matrices 
for $i=1,\ldots,l$
\be\label{taun}
 \tau^{(i)} (\lambda) = 
 \tr_a \Bigl( 
 R_{a,N}^{S_i S_l}(\lambda) \, 
   R_{a,N-1}^{S_i S_{l-1}}(\lambda) \ldots
 R_{a,l+1}^{S_i S_{1}}(\lambda) \, 
 R_{a,l}^{S_i S_l}(\lambda) \, \ldots \,
 R_{a,2}^{S_i S_2}(\lambda) \,
   R_{a,1}^{S_i S_1}(\lambda)  \Bigr) \,.
\ee
They commute with each other and, extending the proof given 
in Section 2, it can be shown that
 \hbox{$\prod_{i=1}^l \tau^{(i)} (0)$} generates a shift 
by~$l$ lattice sites (see Appendix~B). This means that the 
Hamiltonian constructed as follows
\be
 {\cal H}= {\mathrm i}{J \over 2l} \sum_{i=1}^l  \partial_\lambda 
 \ln \tau^{(i)} (\lambda) \Bigm|_{\lambda=0} 
\ee
defines (in the gapless regime) at zero temperature a 
conformally invariant system. We remark that ${\cal H}$ is 
local but involves now interactions of $(l{+}1)$ nearest sites.

The Bethe ansatz equations now read
\begin{equation}\label{BEzn}
 \prod_{j=1}^l
 \left( \frac{ \sinh\gamma(\lambda_\alpha + {\mathrm i}S_j)}%
  { \sinh\gamma(\lambda_\alpha - {\mathrm i} S_j)}  \right)^{N/l}
 = - \prod_{\beta=1}^M 
 \frac{ \sinh\gamma(\lambda_\alpha - \lambda_\beta + {\mathrm i})}%
  { \sinh\gamma(\lambda_\alpha - 
      \lambda_\beta - {\mathrm i}) } \,, \qquad
 \alpha =1,\ldots,M.
\end{equation}
Denote $\q_i=2S_i$. For the Bethe ansatz analysis, the order of 
the spins on the lattice is irrelevant, so we assume in the
following that $\q_1< \q_2 < \ldots < \q_l < \nu$. 
Then, at $T=0$, $H=0$, the ground state in the thermodynamic 
limit is given by 
$\rho_k(\lambda) = {1 \over l} s(\lambda) \delta_{k \q_i}$
(the Dirac sea of $l$ different strings) with the 
corresponding energy 
\be 
 e_{0} = - {J \pi\over l^2} \sum_{i,j=1}^{l}
 \int\limits_{-\infty}^{\infty}d\lambda \,
 Z_{\q_{i}\q_{j}}(\lambda) \, s(\lambda) 
 = \frac{J}{2 l^2} \sum_{i,j=1}^{l} \Bigl[
 \Psi_{q^2} \Bigl( \frac{S_i + S_j}{2} + \frac 12 \Bigr)
 - \Psi_{q^2} \Bigl( \frac{|S_i - S_j|}{2} + \frac 12 \Bigr) \Bigr] . 
\label{energygrn} 
\ee
At $T>0$ the free energy at equilibrium is given by
\be
 f = e_0 - {T\over l} \sum_{i=1}^l \Bigl\{
 \int_{-\infty}^{\infty} d \lambda \, s(\lambda)
 \ln(1+\eta_{\q_{i}}(\lambda))  +
 \delta_{\q_i,\nu-1}  \int_{-\infty}^{\infty} d \lambda \,
 s(\lambda)\ln(1+\eta^{-1}_{0}(\lambda)) \Bigr\} \,.
\label{fn}
\ee 

Equations (\ref{TBA2}) for the pseudo-energies are modified 
only in their free terms by the replacement of 
${1 \over 2} \sum_{i=1}^2 \delta_{k \q_i}$ with 
${1 \over l} \sum_{i=1}^l \delta_{k \q_i}$.
Observe that this change does not affect the 
$\lambda \to \infty$ analysis in the Section 4.3.
For $\lambda \to 0$ we have now $f_{\q_i}=1$ for
$i=1,{\ldots},l$. That is, removing $l$ nodes, we obtain 
$A_n$ diagrams with $n_1=\q_1{-}1$, $n_2=\q_2{-}\q_1{-1}$,
etc. This yields the following central charge
\be
 c(\nu|\q_1,\ldots,\q_l) =  {3\q_{1} \over \q_{1} +2} 
 + \sum_{i=1}^{l-1} {3 \delta \q_i \over \delta \q_i +2} \,,
 \qquad \delta \q_i \equiv \q_{i+1} - \q_i \,.
\label{centraln}
\ee
Notice that the condition $\q_i < \nu$ implies that
$l\leq \nu{-}1$. It is easy to show that, for a given $\nu$,
the maximum of (\ref{centraln}) is attained if 
$l=\nu{-}1$ (that is, if all the possible spins are present),
\be
 \max c(\nu |\{\q_i\}) = c(\nu | 1,2,\ldots,\nu -1) = \nu - 1 \,.
\ee
On the other hand, if $l$ is fixed, then the configuration
$\q_i = i$ corresponds to the minimal possible value,~$c=l$.

The analysis carried out in Section 4.2 extends to the
case of $l$ spins by considering the \hbox{$l$-component}
counterpart of the Wiener--Hopf system~(\ref{system}) (with 
$\xi=\nu/(\nu-\q_l)$ and the coefficients in front of 
$e^{-\pi\lambda}$ being~$k_i/l$).
Then analogues of (\ref{solution}) and (\ref{2}) hold
(with a factor $2/l$ appearing on the r.h.s.~of~(\ref{2})),
and we derive $k^{t} \hat S({\mathrm i}\pi)= 
 {l \over 8 \pi} \bigl(G_{+}(0)G_{-}(0) \bigr)_{ll}$.
Observing that $\bigl(G_{+}(0)G_{-}(0) \bigr)_{ll} = 
 \hat{A}(0)_{\q_l \q_l}$, we obtain 
$k^{t} \hat S({\mathrm i}\pi)= {l \over 4 \pi} {\q_l \over \xi}$.
Using now (\ref{fn}) and proceeding as 
in (\ref{111}), we conclude that
\be 
 f= e_{0} - {1 \over 2 \pi}{\nu \q_{l} \over \nu -\q_{l}}H^{2}\,.
 \label{free1n}
\ee
Whence we infer that
\be
 v_s \, \chi = {1\over 2\pi}{\nu \q_{l} \over \nu -\q_{l}} \,,
\ee
that is, the magnetic susceptibility $\chi$ depends,
for a given anisotropy, only on the maximal spin~$S_l$.

\section{Conclusion}
In this article the alternating quantum spin chain model with 
periodic boundary conditions and its $l$-spin generalization
were investigated via the finite size effects and the algebraic 
Bethe ansatz method. In a previous work \cite{doikou2} the 
boundary free energy for the alternating chain was computed by 
means of the thermodynamic Bethe ansatz method. 
On the other hand, the effective central charge of the open XXZ 
spin chain with general diagonal and non-diagonal boundaries was 
computed recently by means of finite size corrections~\cite{ane}. 
A similar computation, but for diagonal boundaries only, was 
also undertaken in \cite{lmss}. In both cases the effective 
central charge clearly depends on the boundary parameters of 
the model. Let us also mention that the finite size correction 
approach, when boundaries are present, provides somehow the same 
information with the TBA description in the so called R channel 
\cite{cosa}, whereas the thermodynamic Bethe ansatz approach see, 
e.g., \cite{lmss,tsv,doikou2}, can be compared with the TBA 
description in the L channel \cite{lmss,fend}. We hope to 
generalize the finite size correction approach for fused models 
with open boundaries.

To complete the picture, the thermodynamics of the alternating 
spin chain and the $RSOS(\q_{1},\q_{2};q)$ model with integrable 
boundaries should be also investigated. Finally, it would be of 
great interest to continue the thermodynamic investigation for 
spin chains and $RSOS$ models related to higher rank algebras. 
We hope to address these questions in a future work.

{\bf Acknowledgments:} 
We are grateful to A.Babichenko, H.Babujian, V.Tarasov, 
and J.Teschner for useful comments. In particular, we thank 
V.Tarasov for pointing out the reference \cite{TV} to us.
A.B. was supported in part by the Alexander von Humboldt 
Foundation and in part by the Russian Foundation for
Basic Research under grant RFBR-03-01-00593. 
A.D. was supported by the European Network ``EUCLID''; 
``Integrable models and applications: from strings to 
condensed matter'', contract number HPRN-CT-2002-00325.

\section*{Appendix A: universal XXZ R-matrix}
The universal R-matrix (\ref{R1}) was given (in a slightly
different form) in \cite{TV} without an explicit derivation.
We find it instructive to provide here a proof of (\ref{R1}).
Technically, it extends the derivation used in the case 
$S_1=S_2$ \cite{Jim}  (see also \cite{Byt,Tar,Fad1}).
We will also prove some useful related formulae.

Recall that the co-multiplication on $U_q(sl_2)$ is
defined as a linear homomorphism such that
\begin{equation}\label{del}
 \Delta(S^\pm) = S^\pm \otimes q^{-S^3} + q^{S^3} \otimes S^\pm \,,
 \qquad \Delta(S^3) = S^3 \otimes 1 + 1 \otimes S^3 \,.
\end{equation}
Evaluating (\ref{YBE}) in the representation
$V_{\frac 12} \otimes V_{S_1} \otimes V_{S_2}$ 
(and interchanging $\lambda$ with $\mu$), we obtain
\begin{equation}\label{rLL}
 R_{12}^{S_1 S_2}(\lambda) \,  
 L^{S_2}_2 (\lambda+\mu) \, L^{S_1}_1 (\mu) =
 L^{S_1}_1 (\mu) \, L^{S_2}_2 (\lambda+\mu) \,
 R_{12}^{S_1 S_2}(\lambda) 
\end{equation}
which imposes the following conditions on $R^{S_1 S_2}(\lambda)$:
\begin{eqnarray}
 \label{RD}
 R(\lambda) \, \Delta^\prime &=& \Delta \, R(\lambda) \,, \\
 \label{RSK}
  R(\lambda) \, (e^{2 \epsilon \gamma\lambda} 
 S^\epsilon \otimes q^{-S^3} + q^{S^3} \otimes S^\epsilon)  &=&
 (e^{2 \epsilon \gamma\lambda} S^\epsilon \otimes q^{S^3} + 
 q^{-S^3} \otimes S^\epsilon) \, R(\lambda) \,,
\end{eqnarray}
here (and below) $\epsilon = \pm$, and $\Delta^\prime$ is the 
opposite co-multiplication, i.e., 
$\Delta^\prime_q = \Delta_{1/q}$. 
These equations determine the universal R-matrix uniquely (up to 
a scalar factor) \cite{Jim,TV}. Denote $R_\pm = (R(\pm\infty))^{-1}$. 
Then the asymptotics of (\ref{RD})--(\ref{RSK}) yields
\begin{eqnarray}
 \label{Di}
  R_\epsilon \, \Delta &=& \Delta^\prime \, R_\epsilon \,, \\
 \label{Ri}
 R_\epsilon \, ( S^\epsilon \otimes q^{S^3} ) =
 ( S^\epsilon \otimes q^{-S^3} ) \, R_\epsilon \,, &&
 R_\epsilon \, ( q^{-S^3} \otimes S^{-\epsilon} ) =
 ( q^{S^3} \otimes S^{-\epsilon} ) \, R_\epsilon \,.
\end{eqnarray}
These equations imply that $R_\pm$ are the constant 
universal R-matrices~(\ref{Rc}). Observe that (\ref{Di}) 
implies that $R_\epsilon$ maps the $\Delta$--basis in 
$V_{S_1} \otimes V_{S_2}$ into the $\Delta^\prime$--basis. 
That is, for the orthonormal set of eigenvectors 
$|j,m \rangle$, $m=j,j-1,...,-j$ of a projector ${\cal P}_j$
we have
\be\label{Rj}
 R^{S_1 S_2}_\epsilon |j,m \rangle = \varphi^\epsilon_j \, 
 |j,m \rangle^\prime \,, 
\ee 
where the scalar factor $\varphi^\epsilon_j$ does not depend on~$m$
(its explicit form is given below).

Now we introduce $r^\pm(\lambda) = R(\lambda) R_\pm$ and then
(\ref{RD})--(\ref{RSK}) acquire the form
\begin{eqnarray}
 \label{rD}
  r^\epsilon (\lambda) \, \Delta &=& 
    \Delta \, r^\epsilon (\lambda) \,, \\
 \label{rSK}
  r^\epsilon(\lambda) \, \bigl( e^{2 \epsilon \gamma\lambda} 
     S^\epsilon \otimes q^{S^3} + 
 (R_\epsilon)^{-1} \bigl( q^{S^3} \otimes S^\epsilon \bigr) 
 R_\epsilon \bigr)  &=& 
 \bigl( e^{2 \epsilon \gamma\lambda} S^\epsilon \otimes q^{S^3} + 
 q^{-S^3} \otimes S^\epsilon \bigr) \, r^\epsilon(\lambda) \,.
\end{eqnarray}
It follows from (\ref{rD}) that $r^\pm(\lambda)$ are
functions of the spin operator $\mathbb J$, i.e.,
\be\label{rP}
  r^\pm(\lambda) = \sum_{|S_1-S_2|}^{S_1+S_2} 
  r^\pm_j (\lambda) \, {\cal P}_j \,,
\ee
where $r^\pm_j (\lambda)$ are scalar functions and ${\cal P}_j$ 
is the projector onto $V_j$ in~$V_{S_1} \otimes V_{S_2}$.

Using (\ref{del}), (\ref{Ri}), and (\ref{Rj}), it is 
straightforward to check the following relations
\begin{eqnarray}
 \label{eps1}
&& [ S^\epsilon \otimes q^{S^3}, \Delta(S^\epsilon) ] =
 [ q^{-S^3} \otimes S^\epsilon , \Delta(S^\epsilon) ] = 
 [ (R_\epsilon)^{-1} (q^{S^3} \otimes S^\epsilon) R_\epsilon ,
 \Delta(S^\epsilon) ] = 0 \,, \\ [0.5mm]
 \label{eps2}
&& ( S^\epsilon \otimes q^{S^3} + q^{\epsilon (2+2j)}
 q^{-S^3} \otimes S^\epsilon) \, |j,\epsilon j \rangle =
 (1\otimes q^{2S^3}) \, \Delta (S^\epsilon) \, 
 |j, \epsilon j \rangle = 0 \,,  \\ [0.5mm]
 \label{eps3}
&& \Bigl(  \bigl( q^{S^3} \otimes S^\epsilon \bigr) R_\epsilon + 
 q^{\epsilon (2+2j)} R_\epsilon 
 \bigl( S^\epsilon \otimes q^{S^3} \bigr) \Bigr) \, 
 |j,\epsilon j \rangle = (q^{2S^3} \otimes 1) \, 
 \Delta^\prime (S^\epsilon) \, |j, \epsilon j \rangle^\prime = 0 \,.
\end{eqnarray}

Eqs.~(\ref{rP})--(\ref{eps1}) show that we can apply (\ref{rSK})  
to a highest/lowest weight vector $|j,\epsilon j \rangle$.
Doing so and taking (\ref{eps2})--(\ref{eps3}) into account,
we obtain
\be\label{rjja}
  r^\epsilon_{j+1}(\lambda) \, (e^{ \epsilon \gamma\lambda} -
 q^{\epsilon (2+2j)} e^{- \epsilon \gamma\lambda}) =
 (e^{\epsilon \gamma\lambda} - 
    q^{-\epsilon(2+2j)} e^{-\epsilon \gamma\lambda})
 \ r^\epsilon_{j}(\lambda) \,. 
\ee
Hence 
\begin{equation}\label{rj}
 r^\epsilon_{j+1}(\lambda) = - q^{-\epsilon (2+2j)} \,
 \frac{ [j+1-{\mathrm i}\lambda]_q}{[j+1+{\mathrm i}\lambda]_q}  
  \  r^\epsilon_{j}(\lambda)  \,.
\end{equation}
Solving this functional equation and fixing the overall 
normalization, we obtain (\ref{R1}).

Notice that (\ref{R1}) has the form 
$q^{\psi_\epsilon({\mathbb J})} \Omega(\lambda,{\mathbb J}) 
 \bigl(R_\epsilon)^{-1}$, where the functions
$\psi_\epsilon(x)$ and $\Omega(\lambda,x)$ are invariant
with respect to the replacement of $q$ by $q^{-1}$,
and, moreover, $\Omega(\lambda,x)\Omega(-\lambda,x)=1$,
$\psi_{-\epsilon}(x)=-\psi_\epsilon(x)$.
Whence, taking into account the property 
$(R_\epsilon)_{1/q} = (R_\epsilon)^{-1}$, we 
derive the first symmetry relation in~(\ref{Rsym})
\be
 R_{1/q}(-\lambda) =  
 q^{-\psi_\epsilon({\mathbb J}^\prime)} 
 \Omega(-\lambda,{\mathbb J}^\prime) \, R_{\epsilon}   = 
 R_\epsilon \, q^{-\psi_\epsilon({\mathbb J})} 
 \bigl( \Omega(\lambda,{\mathbb J}) \bigr)^{-1} =
 \bigl( R_q(\lambda) \bigr)^{-1} \,.
\ee
Here ${\mathbb J}^\prime = {\mathbb J}_{1/q}$
is the two--node spin operator corresponding to the 
opposite co-multiplication and we used that 
$R_\epsilon {\mathbb J} = {\mathbb J}^\prime 
 R_\epsilon$, as follows from~(\ref{Di}).

If the algebra generators are represented in terms of
matrices as in (\ref{sss}), then it is easy to check that 
${\mathbb J}^t={\mathbb J}$ and $(R_+)^t = (R_-)^{-1}$,
where $t$ stands for the matrix transposition. Whence
we derive the second symmetry relation in~(\ref{Rsym}),
\be
 \bigl( R(-\lambda) \bigr)^t = R_{-\epsilon} \,
 q^{\psi_\epsilon({\mathbb J})} 
 \Omega(-\lambda,{\mathbb J})  = 
 R_{-\epsilon} \, q^{-\psi_{-\epsilon}({\mathbb J})} 
 \bigl( \Omega(\lambda,{\mathbb J}) \bigr)^{-1} =
 \bigl( R_q(\lambda) \bigr)^{-1} \,.
\ee

It is straightforward to see from (\ref{Pj})-(\ref{dC2}) 
that action of the $*$-operation (\ref{star}) on the
two--node spin operator ${\mathbb J}$ is given 
by~$({\mathbb J}_q)^* = {\mathbb J}_{\bar{q}}$. This, along 
with the property $(R_{\pm,q})^*=(R_{\mp,\bar{q}})^{-1}$,
leads to the first relation in (\ref{*R}):
\be\label{R*}
 \bigl( R(\lambda) \bigr)^* = R_{-\epsilon, \bar{q}} \,
 \Omega(-\bar{\lambda}, {\mathbb J}_{\bar{q}}) \,
 \bar{q}^{\,-\psi_{-\epsilon}({\mathbb J}_{\bar{q}} )} 
 = \bigl( R_{\bar{q}}(\bar{\lambda}) \bigr)^{-1} \,.
\ee
The second relation in (\ref{*R}) follows from (\ref{Rh}),
(\ref{R*}), and the symmetries (\ref{Rsym}) and~(\ref{Rhsym}).
For instance, for $\gamma\in {\mathbb R}$ we have
\begin{eqnarray}
 \nonumber  \bigl( \hat{R}_q(\lambda) \bigr)^* &=& 
 e^{ - \gamma \bar{\lambda} (1\otimes S^3) } \, 
 \bigl( R_{1/q}(\bar{\lambda}) \bigr)^{-1} \,
 e^{ \gamma \bar{\lambda} (1\otimes S^3) }  \\
 &=& e^{ - \gamma \bar{\lambda} (1\otimes S^3) } \, 
 R_{q}(-\bar{\lambda}) \,
 e^{ \gamma \bar{\lambda} (1\otimes S^3) } =
 \hat{R}_q(-\bar{\lambda}) =
 \bigl( \hat{R}_q(\bar{\lambda}) \bigr)^{-1} .
\end{eqnarray}

To prove the important identity (\ref{JP}), we employ  
Theorem~3.1 from~\cite{KR2} that states (in our notations) 
that the $U_q(sl_2)$ Clebsch-Gordan coefficients satisfy
the following relations
\begin{eqnarray}
\label{KR1}
  \sum_{m_1^\prime,m_2^\prime} 
 \bigl(R^{S_1 S_2}_+\bigr)^{m_1^\prime,m_2^\prime}_{m_1,m_2} \,
 \Bigl[ {S_1 \atop m_1}\, {S_2 \atop m_2}\, 
 {j \atop m} \Bigr]_q  &=& 
 (-1)^{S_1+S_2 -j} q^{ \kappa(S_1) + \kappa(S_2) - \kappa(j)} \,
 \Bigl[ {S_2 \atop m_2}\, {S_1 \atop m_1}\, 
 {j \atop m} \Bigr]_q  , \\
\label{KR2}
  \Bigl[ {S_1 \atop m_1}\, {S_2 \atop m_2}\, 
 {j \atop m} \Bigr]_q &=& 
 (-1)^{S_1+S_2 -j} \Bigl[ {S_2 \atop m_2}\, {S_1 \atop m_1}\, 
 {j \atop m} \Bigr]_{q^{-1}} ,
\end{eqnarray}
where $\kappa(x) \equiv x(x+1)$, and
$\bigl(R^{S_1 S_2}_+\bigr)^{m_1^\prime,m_2^\prime}_{m_1,m_2}
 = \langle m_1^\prime | \otimes \langle m_2^\prime |
 R^{S_1 S_2}_+ |m_1\rangle \otimes |m_2\rangle$
are matrix elements of the R-matrix in the 
$V_{S_1}\otimes V_{S_2}$ basis.
These formulae show that for (\ref{Rj}) we have
\be\label{phip}
 \varphi^+_j = q^{ \kappa(S_1) + \kappa(S_2) - \kappa(j)  }  \,,  
\ee
and also that 
\be
 \langle m_1^\prime | \otimes \langle m_2^\prime | \,
 R^{S S}_+ \, |j,m \rangle =  (-1)^{2S-j} \varphi^+_j 
 \langle m_1^\prime | \otimes \langle m_2^\prime | \,
 {\mathbb P} \, |j,m \rangle \,,
\ee 
which proves (\ref{JP}) for $\epsilon=+$. Analogues formulae
for $\epsilon=-$ are obtained by observing that the map 
$S^+ \leftrightarrow S^-$, $S^3 \leftrightarrow -S^3$, 
$q \leftrightarrow q^{-1}$ is an automorphism preserving 
the co-multiplication~(\ref{del}). In particular, we have
\be
\varphi^-_j=(\varphi^+_j)^{-1}
\ee
which along with (\ref{Rj}) and (\ref{phip}) leads to the
following relations
\be\label{RRP}
 R_-^{-1} R_+ = \sum_{j=\delta S}^{S_1 + S_2}
  (\varphi^+_j)^2 {\cal P}_j \,, \qquad
 R_+^{-1} R_- = \sum_{j=\delta S}^{S_1 + S_2}
  (\varphi^-_j)^2 {\cal P}_j \,.
\ee
Inverting these relations, we obtain 
\be\label{PRR0}
 {\cal P}_j = \prod_{ {l=\delta S} \atop {l\neq j} }^{S_1 + S_2} \, 
 \frac{ (\varphi^\epsilon_l)^2 - R_{-\epsilon}^{-1} R_\epsilon }%
 { (\varphi^\epsilon_l)^2 - (\varphi^\epsilon_j)^2 } \,, \qquad
 \varphi^\epsilon_j = q^{ \epsilon ( \kappa(S_1) + 
 \kappa(S_2) - \kappa(j) ) } \,.
\ee
Indeed, to prove this formula, it suffices first to observe that, 
as seen from (\ref{RRP}), the only projector which is present 
in all the factors of the r.h.s.~of (\ref{PRR0}) is~${\cal P}_j$,
and then to check the corresponding normalization.
In the simplest case, $(S_1,S_2)=(\frac 12,S)$, 
eq.~(\ref{PRR0}) yields
\be
 {\cal P}_{S \pm \frac 12}^{\frac{1}{2} S} =
 \pm \frac{1}{[2S+1]}_q \, \Bigl( q^{\mp (2S+1)} - 
 q\, \bigl( R_{+}^{\frac{1}{2} S} \bigr)^{-1} 
 R_-^{\frac{1}{2} S} \Bigr) \,.
\ee

Consider now the universal R-matrix introduced in (\ref{Rh}).
It is a solution to equation (\ref{rLL}) for the $L$-operator 
(\ref{Lh}). The latter, unlike the L-operator (\ref{Lop}), 
does not decompose into two Borel components. As a consequence,
instead of (\ref{RD})-(\ref{RSK}), we have 
four $\lambda$-dependent relations:
\begin{eqnarray}
 \label{rhSa}
 && \hat{R}(\lambda) \, \bigl( e^{\epsilon \gamma\lambda}
 S^\epsilon \otimes q^{-S^3} +
   q^{S^3} \otimes S^\epsilon \bigr) =
 \bigl( e^{\epsilon \gamma\lambda} 
 S^\epsilon \otimes q^{S^3} + q^{-S^3} \otimes S^\epsilon 
 \bigr) \, \hat{R}(\lambda) \,, \\ 
 \label{rhSb}
 && \hat{R}(\lambda) \, \bigl( e^{\epsilon \gamma\lambda}
 q^{-S^3} \otimes S^\epsilon +
 S^\epsilon \otimes q^{S^3} \bigr) =
 \bigl( e^{\epsilon \gamma\lambda} 
 q^{S^3} \otimes S^\epsilon + 
 S^\epsilon \otimes q^{-S^3} \bigr) \, \hat{R}(\lambda) \,. 
\end{eqnarray}
Observe that $\tilde{R}(\lambda) \equiv (\hat{R}(\lambda))^t$
solves the same set of equations, which, taking into account that 
the solution to (\ref{rhSa})-(\ref{rhSb}) is unique (up to a
scalar factor), implies that 
$(\hat{R}(\lambda))^t = \zeta(\lambda) \hat{R}(\lambda)$,
with $\zeta(\lambda)$  being a scalar function. In fact,
$\zeta(\lambda) = \pm 1$ because $t \circ t = id$. The 
possibility of $\zeta(\lambda) = - 1$ is excluded by the fact 
that the vector $|S_1\rangle \otimes |S_2\rangle$ is an
eigenvector of $\hat{R}^{S_1 S_2}(\lambda)$ with a non-vanishing
eigenvalue (hence the first matrix element of
$\hat{R}^{S_1 S_2}(\lambda)$ in the basis corresponding to (\ref{sss})
is non-vanishing). Thus, we infer that $\zeta(\lambda) = 1$,
which proves the last relation in (\ref{Rhsym}). The remaining 
symmetries in (\ref{Rhsym}) are easily derived from this 
symmetry and the above established relations (\ref{Rsym}),
for instance
\begin{eqnarray}
 \hat{R}(-\lambda) &=& e^{ -\gamma \lambda (1\otimes S^3) } \, 
 R(-\lambda) \, e^{ \gamma \lambda (1\otimes S^3) } \\
 \nonumber  &=& 
 \bigl( e^{ -\gamma \lambda (1\otimes S^3) } \, 
 (R(\lambda))^t \, e^{ \gamma \lambda (1\otimes S^3) } \bigr)^{-1} 
 = \bigl( (\hat{R}(\lambda))^t \bigr)^{-1} = 
 \bigl( \hat{R}(\lambda) \bigr)^{-1} \,.
\end{eqnarray}
Finally, since $\hat{R}(0) = R(0)$, the symmetries (\ref{R0sym}) 
follow obviously from (\ref{Rsym}) and (\ref{Rhsym}).

\section*{Appendix B: Construction of the shift operator}
Let $l$ and $N$ be integer such that $l$ divides~$N$.
Introduce the following invertible operators 
\be
 Q^{i,j}_\pm = \prod_{k= i \bmod l}^N 
    R^{S_i\, S_j}_{k,k+j-i \pm l}(0) \,, \qquad
 \tilde{Q}^{i,j} = \prod_{k=i \bmod l}^N 
    R^{S_i\, S_j}_{k,k+j-i}(0) \,,
\ee
where 
$R^{S_i\, S_j}_{n,m}(\lambda)$ is the XXZ R-matrix (assigned to 
the sites $n$ and $m$ of the lattice in the sense of Section~2).
Notice that in each product all the factors  commute with
each other iff~$i \neq j \bmod l$. Along with the property 
(\ref{RR0}) this implies that
\be
 Q^{i,j}_+ \, Q^{j,i}_- = Q^{i,j}_- \,  Q^{j,i}_+
 = 1^{\otimes N} \,.
\label{QQ}
\ee
Consider now the transfer matrices (\ref{taun}). It is not
difficult to derive for them an analogue of eq.~(\ref{tau0}),
\be
 \tau^{(i)} (0) = U^{(i)} \, Q^{i,i-1}_+ \, Q^{i,i-2}_+ \ldots
 Q^{i,1}_+ \, \tilde{Q}^{i,l} \, \tilde{Q}^{i,l-1} 
 \ldots  \tilde{Q}^{i,i+1} \,,  
\label{tQ}
\ee 
where $U^{(i)} = \prod_{k=i \bmod l}^N {\mathbb P}_{k,k+l}$
(the factors are ordered from the left to the right)
acts nontrivially only on the nodes $n = i \bmod l$, 
\be\label{Uxn}
 U^{(i)} \, \xi_n = \xi_{n+l} \, U^{(i)} \qquad
 {\rm if}\ \  n\equiv i \bmod l \,.
\ee
Observe that $U=U^{(1)} U^{(2)} \ldots U^{(l)}$ is the lattice
shift operator (generating a shift by~$l$ sites).

Notice that $\tilde{Q}^{i,j} U^{(j)} = U^{(j)} Q^{i,j}_-$. 
Therefore (taking all $U^{(j)}$ to the left) we have
\be 
 \tau^{(1)} (0) \, \tau^{(2)} (0) \ldots \tau^{(l)} (0) = 
 & U & Q^{1,l}_- \, Q^{1,l-1}_- \ldots Q^{1,3}_- \, Q^{1,2}_-  \\
 &\times& Q^{2,1}_+ \, Q^{2,l}_- \ldots Q^{2,4}_- \, Q^{2,3}_- \non \\
 && \qquad \qquad \cdots  \non \\
 &\times& Q^{l-1,l-2}_+ \, Q^{l-1,l-3}_+ 
       \ldots Q^{l-1,1}_+ \, Q^{l-1,l}_- \non \\
 &\times& Q^{l,l-1}_+ \, Q^{l,l-2}_+ 
   \ldots Q^{l,2}_+ \, Q^{l,1}_+ \non  \,.
\ee
Notice that here we have $Q^{i,j}_-$ if $i<j$ and $Q^{i,j}_+$
if $i>j$. Let us show that all $Q_-$'s can be led to meet their
$Q_+$ counterparts and thus be canceled by the relation~(\ref{QQ}).
As the first step, the rightmost $Q_-$ in each row, i.e, $Q^{i,i+1}_-$, 
$i=1,\ldots,l{-}1$, is canceled against the leftmost $Q_+$
in the next row. Observe that this eliminates obstacles for moving 
then the second from the right $Q_-$ through the entire next row
(obviously, $Q^{i,j}_\pm$ commutes with $Q^{i',j'}_\pm$
if $i,j,i',j'$ are all distinct). Thus, at the second step,
$Q^{i,i+2}_-$ for $i=1,\ldots,l{-}2$ are canceled against
$Q^{i+2,i}_+$ (which, after the first step, became the 
leftmost factors in their rows). Proceeding this way, we
eliminate at the $k$-th step $l{-}k$ pairs 
$(Q^{i,i+k}_-, Q^{i+k,i}_+)$ and also make possible the next step 
(that is, $Q^{i,i+k+1}_-$ will be able to move through the next $k$ 
rows). Consequently, after the $(l{-}1)$-th step we will have 
all $Q$'s eliminated, and so we conclude that 
$\tau^{(1)} (0)  \ldots \tau^{(l)} (0) = U$.

\end{document}